\documentclass{article}

\usepackage{PRIMEarxiv}

\usepackage[utf8]{inputenc} 
\usepackage[T1]{fontenc}    
\usepackage{hyperref}       
\usepackage{url}            
\usepackage{booktabs}       
\usepackage{amsfonts}       
\usepackage{nicefrac}       
\usepackage{microtype}      
\usepackage{lipsum}
\usepackage{graphicx}
\graphicspath{{media/}}     
\usepackage{appendix} 
\usepackage{tikz}
\usepackage{amsmath}
\usepackage{amssymb} 
\usepackage{booktabs} 
\usepackage{tabularx} 
\usepackage{array}
\usepackage{listings}
\title{Can AI Lower the Barrier to Cybersecurity? \\A Human-Centered Mixed-Methods Study of Novice CTF Learning}

\author{
Cathrin Schachner\thanks{In alphabetical order. Both authors contributed equally to this work.}\\
  University of Klagenfurt \\
  Klagenfurt\\
  \AND
  Jasmin Wachter \thanks{Corresponding Author. Both authors contributed equally to this work.} \\
  University of Klagenfurt \\
  Klagenfurt\\
  \texttt{Jasmin.Wachter@aau.at} \\
}

\begin{document}
\maketitle

\begin{abstract}
Capture-the-Flag (CTF) competitions serve as  gateways into offensive cybersecurity, yet they often present steep barriers for novices due to complex toolchains and opaque workflows. Recently, agentic AI frameworks for cybersecurity promise to lower these barriers by automating and coordinating penetration testing tasks. However, their role in shaping novice learning remains underexplored.

We present a human-centered, mixed-methods case study examining how agentic AI frameworks -- here Cybersecurity AI (CAI)\cite{mayoralvilches2025caiopenbugbountyready} -- mediates novice entry into CTF-based penetration testing. An undergraduate student without prior hacking experience attempted to approach performance benchmarks from a national cybersecurity challenge using CAI. Quantitative performance metrics were complemented by structured reflective analysis of learning progression and AI interaction patterns.

Our thematic analysis suggest that agentic AI reduces initial entry barriers by providing overview, structure and guidance, thereby lowering the cognitive workload during early engagement. Quantitatively, the observed extensive exploration of strategies and low per-strategy execution time potetially facilitatates cybersecurity training on meta, i.e. strategic levels. At the same time, AI-assisted cybersecurity education introduces new challenges related to trust, dependency, and responsible use. We discuss implications for human-centered AI-supported cybersecurity education and outline open questions for future research.
\end{abstract}


\section{Introduction}

Capture-the-Flag (CTF) competitions are widely used in cybersecurity education \cite{SVABENSKY2021102154} and talent identification. They provide hands-on exposure to penetration testing techniques, vulnerability discovery, and adversarial thinking. Despite their pedagogical value, CTF environments often present steep entry barriers for novices. CTF-Challengers are expected to navigate complex toolchains, interpret unfamiliar outputs, and strategically coordinate multiple attack steps -- skills typically acquired through prior experience. For individuals without prior exposure to penetration testing, this initial barrier may discourage participation altogether. Lowering this threshold without compromising meaningful skill development remains an open challenge in cybersecurity education.

Recent advances in large language models (LLMs) and agentic frameworks have enabled semi-automated orchestration of cybersecurity workflows. Frameworks such as Cybersecurity AI (CAI) integrate language models with traditional penetration testing tools, supporting reconnaissance, vulnerability analysis, and report generation.

Proponents argue that such systems may democratize access to offensive security by reducing technical friction and increasing transparency of complex workflows. At the same time, concerns remain regarding over-automation, skill atrophy, ethical misuse, and shifting trust dynamics. While existing benchmarking efforts primarily evaluate AI performance in technical tasks, less is known about how these systems shape human learning and novice participation.

This work takes a didactic, human-centered perspective on cybersecurity AI. The paper addresses the question \textit{whether agentic systems function merely as automation tools -- or as facilitators of early-stage learning}. 

To explore this question, we investigate the following research questions:
\begin{itemize}
    \item[\textbf{RQ1:}] To what extent does AI assistance enable a novice learner without prior penetration testing experience to approach performance benchmarks established by CTF-attempters of a national cybersecurity challenge?
    \item[\textbf{RQ2:}] How does an agentic AI framework mediate novice entry into penetration testing practice?
\end{itemize}
Rather than treating AI performance as the primary outcome, we focus on how AI reshapes engagement, strategy formation, confidence, and learning processes. Our contributions are as follows:
\begin{itemize}
    \item \textbf{Logitudinal empirical study of agentic AI-mediated novice entry into CTF practice}.
	\item \textbf{Human-centered evidence that agentic AI lowers cognitive entry barriers through structured orientation}.
	\item \textbf{Quantitative indications of strategic acceleration under (C)AI assistance.}
	\item \textbf{Implications for responsible integration of agentic AI into cybersecurity training.}
\end{itemize}
\section{Related Work and Background}
\paragraph{Agentic AI in Offensive Cybersecurity}
The integration of large language models (LLMs) into cybersecurity workflows has led to the development of agentic AI frameworks capable of coordinating reconnaissance, vulnerability analysis, and tool orchestration. Early work explored integrating LLMs into penetration testing pipelines and security test suites \cite{happe2023getting}; others PentestGPT \cite{deng2024pentestgpt} demonstrate the use of LLMs for structured offensive workflows. More recently, semi-autonomous, agentic frameworks, such as Cybersecurity AI (CAI) \cite{mayoralvilches2025caiopenbugbountyready}, HackingBuddyGPT \cite{happe2025llms}, HexStrike AI MCPs \cite{one7l2026hexstrike} and related orchestration platforms, combine LLM reasoning with established penetesting tools calling.

The emerge of these systems shifts the role of AI in security practise from isolated assistance towards integrated and coordinated multi-step workflows. Moreover, the tools became increasingly accessible, and questions arise not only about their technical capabilities but also about their impact on cybersecurity education and entry pathways.

\paragraph{Benchmarking AI performance in Security Tasks}
 A growing body of research evaluates the technical performance of AI systems in offensive security contexts. Benchmarking approaches \cite{happe2025benchmarking} include structured security quizzes and task inventories \cite{sanz2025cybersecurity, zhang2024cybench}, red-team simulations, and comparisons with human experts \cite{Petrov2025}. Some studies explore whether AI systems may rival or surpass human pentesters in capability \cite{mayoralvilches2025caiopenbugbountyready}.

While these studies provide important insight into system-level capability, they primarily focus on AI performance in isolation. Less attention has been paid to how AI assistance shapes the learning processes, strategic behavior, and participation trajectories of novice users.
In contrast to system-level benchmarking studies, this work adopts a human-centered perspective on agentic cybersecurity AI. Rather than asking whether AI can solve CTF challenges, we examine how AI assistance mediates novice entry into penetration testing practice.

\paragraph{Uneven Entry Pathways into Cybersecurity Education}
Entry into offensive cybersecurity practice is often shaped by prior exposure to technical training. In Austria, secondary education diverges into technical and non-technical tracks, leading to substantial variation in early access to computer science and cybersecurity curricula \cite{OesterreichHTLCurriculum, AustrianComputerScienceCurriculum}. Students graduating from technical high schools (HTLs) frequently receive structured exposure to networking and security topics, whereas others may encounter cybersecurity only at the university level\footnote{Notably, evidence confirms that social differences related to gender, migration status, and educational background \cite{steiner2023bildungsverlauf, Fassl2012_WomenTransitionSTEM} disproportionately influence enrollment in non-technical high schools.}. 
Although this study does not aim to quantify structural disparities, it situates the present case within a broader context of uneven preparation for CTF participation\footnote{Many Austrian CTF-teams are in fact school teams from technical high schools}. Such disparities highlight the importance of tools that may reduce entry barriers for individuals with adjacent technical backgrounds but limited exposure to offensive security practice.

\paragraph{AI as a Mediator of Learning and Practice}
Research in human-centered AI and AI-assisted learning emphasizes that intelligent systems do not merely automate tasks but reshape how users engage with complex domains. Studies highlight both the potential of AI systems to support exploration and reduce cognitive friction \cite{riva2025cognitive}, as well as risks related to overreliance and diminished critical evaluation \cite{gerlich2025ai}, cognitive effort and confidence \cite{lee2025impact}. 

Regarding skill formation in computer science, a recent study \cite{shen2026ai}
 suggests that using AI to complete tasks that require a new skill (here: a know software library) reduces skill formation, while only slightly reducing the average time taken (not significant). 

Frameworks of AI fluency \cite{dakan2025ai, mayoralvilches2025caifluencyframeworkcybersecurity} therefore suggest that effective humane-AI interaction requires competencies such as strategic delegation, critical discernment, and responsible use -- also in cybersecurity \cite{mayoralvilches2025caifluencyframeworkcybersecurity}. 

In this context, evaluating AI solely in terms of task performance therefore overlooks its broader role as a mediator of human activity. 

\section{Methodology}
 \label{sec:Methodology}
To examine how AI assistance mediates novice-but-skilled entry into penetration testing practice, we conducted a longitudinal mixed-methods case study combining performance benchmarking with an autoethnography-inspired action research approach, emphasizing the individual and reflexive nature of the study. While not a full autoethnography, the action-research part centers on the experiences of the subject of the research -- one of the authors -- who logged her experiences throughout the process, while the second author analyzed and coded reflections. These logs were recapitulated and indirectly incorporated in the final structural reflection questionnaire that was analyzed using thematic analysis.
 Second, a broader survey group, namely last years' Austria Cybersecurity Challenge (ACSC) Participants and Challengers (i.e. people who simply tried the CTFs off the official challenge), were examined. The study examins both measurable outcomes (relative CTF performance) and experiential dimensions, particularily learning and AI interaction patterns in terms of user confidence and usability.

\subsection{Subjects under Scrutiny}
The \textbf{primary subject} was an undergraduate computer science student with approximately six years of vocational system operating and engineering experience, but no prior experience in CTFs, penetration testing, or cybersecurity defense beyond a single university course. This positioned the participant as a novice entering offensive security practice for the first time, making this study particularly relevant to the design and evaluation of tools and training programs. As an individual with a non-technical high school (HTL) background, the insights gained from this research are informed by a unique perspective within the cybersecurity field.

To contextualize these findings, we additonally surveyed 29 cybersecurity practitioners (aka. the \textit{ACSC CTF-attempters}) from last years Austria Cybersecurity Challenge using a LimeSurvey online questionnaire.  
\paragraph{Recruitment}
The participants all tried the CTFs of last years ACSC. Recruitment occurred via the ACSC Discord, where we shared the link to the lime-survey online questionnaire, stating all participation is anonymous, voluntary, and without compensation. The initial $N=33$ responses reduced to $N=29$ due to incomplete general part in the questionnaire.
\paragraph{Questionnaire}
The research instrument comprised of a general part covering experience, use of AI, and educational background, and a part dedicated to the five ACSC CTF challenges described in table \ref{tab:acsc_ctf_challenges}; cf. Appendix \ref{sec:appendix3} for the full research instrument. 

Note that data collection varied across CTF challenges due to the implementation of filters requiring responses to ``did you try this challenge?'' and ``did you solve this challenge?'' questions. Additionally, participant attrition was observed in the second part of the survey, with some individuals not completing the full study protocol table, cf. \ref{tab:ctf_filter} for the individual sample sizes.
\paragraph{Demographics}Demographically, the ACSC survey participants were largely (self-reported) intermediate practitioners (45\%), with smaller groups identifying as beginners (20,5\%), advanced (20,5\%), and experts (14\% each). Self-study was the most common learning pathway (93\%), followed by formal education (55\%) and on-the-job experience (31\%). While AI adoption is growing, primarily for automation (55\%) via tools, and augmentation (52\%), where individual tasks, such as coding are outsourced to AI (e.g. LLMS), autonomous use using agents remains limited (10\%). Additionally, 31\% claimed to have no experience integrating AI into their workflows.

\subsection{Study Outline}
To address RQ1, the primary study subject was equipped with (but not familiar with) CAI \cite{mayoralvilches2025caiopenbugbountyready}, an agentic LLM-cybersecurity framework. The subject's objective was to replicate or surpass the performance of last year's ACSC participants ($N=29$) on a reconstructed version of the official challenge environment. Unlike the other participants, the primary subject was equipped with \textit{Cybersecurity AI} (CAI) \cite{mayoralvilches2025caiopenbugbountyready}, an open-source framework designed to help ethical hackers and researchers build AI-driven security tools (cf. \textit{vibe-hacking} \cite{mayoralvilches2025caifluencyframeworkcybersecurity}). Both quantitative and qualitative data (see Section \ref{sec:measures}) was collected to evaluate the performance of the tool and the neo-pentester.

For RQ2, an autoethnography-inspired \footnote{``Autoethnographers must not only use their methodological tools and research
literature to analyze experience, but also must consider ways others may experience similar epiphanies; they must use personal experience to illustratefacets of cultural experience, and, in so doing, make characteristics of a culture familiar for insiders and outsiders.'' \cite{ellis2011autoethnography}} \cite{ellis2011autoethnography} action research approach \cite{kock2014actionresearch} was conducted, where the primary subject documented and reflected on personal experiences and progress as an inexperienced pentester, exploring how AI tools facilitated the entry into penetration testing by assessing factors such as functionality, performance and usability \cite{peffers2012design}. This study enabled an in-depth inspection of the learning process
and the role of AI tools in improving cybersecurity skills. The structured retrospective in written from materializes the final reflection process. 

The \textbf{longitudinal study} consisted of four phases (Ph.) over almost a year time:
\begin{enumerate}
\item[\textbf{\small{Ph. 1:}}] \textbf{CAI Introduction \& Setup:} The participant received initial exposure to the agentic Cybersecurity AI (CAI) framework and the challenge environment, while simultaneously documenting their initial experiences in technical logs as well as field notes.
\item[\textbf{\small{Ph. 2:}}] \textbf{Iterative Challenge Solving (Cyberleague):} The participant iteratively solved five challenges from the Cyberleague CTF competition using CAI (see table \ref{tab:cyberleague_challenges}, cf. Appendix \ref{sec:appendix3}), recording their progress and observations.
\item[\textbf{\small{Ph. 3:}}] \textbf{Quantitative Benchmarking (ACSC):} The participant quantitatively benchmarked their performance against ACSC participants on five challenges from last year's CTFs, utilizing CAI (see table \ref{tab:ctf_challenges}, cf. Appendix \ref{sec:appendix3}), and maintained logs. This phase included a jointly developed questionnaire for the ACSC participants, with statistical evaluation and qualitative coding of field notes conducted by both authors.
\item[\textbf{\small{Ph. 4:}}] \textbf{Retrospective Reflection:} In restrospective, the primary subject completed a structured retrospective written reflection on their experiences, based on a structured guide created by the other author, who analyzed it.
\end{enumerate}

\subsection{Materials and Environment}
As mentioned before, the study employed a reconstructed challenge environment modeled on tasks from the \textit{Austria Cybersecurity Challenge (ACSC)} as well as \textit{cyberleague.at}. Moreover, the open-source Cybersecurity AI (CAI) framework \footnote{CAI Community Edition provides a complete open-source framework for research and learning, while CAI PRO offers additional features and support for a monthly fee. As of summer 2025, CAI has diverged into the open-source CAIO branch and a proprietary branch. In this study we utilized the ancestor CAI 0.5 version before the divergence. The functionality is, however, equivalent to the now freely available Community Edition.} was utilized, in combination with LLMs, specifically OpenAI ChatGPT Models and Anthropic Claude Models. The selection of CAI as the agentic cybersecurity tool under scritiny was based on one the other authors familiarity with the tool and the developers. The primary subject was not familiar with the tool before the study.\footnote{Note that we initiated the study at the end of 2024, when benchmarks, open-source tools, and comparative data were largely unavailable.}

Additionally, the integrated tools such as Nmap, Burp Suite, and related penetration testing utilities were employed. 
For performance benchmarks, a survey among ACSC CTF challengers ($N = 29$) was conducted. 

\begin{table}[h]
\centering
\caption{CyberLeague Challenges employed in Ph.2 to target RQ1}
\begin{tabularx}{0.75\textwidth}{p{3cm}X}
\toprule
\textbf{CTF Challenge} & \textbf{Description} \\
\midrule
\textit{All About that Base} (Crypto, Beginner) & A reversing/encoding puzzle. The participant recovers the original flag by reversing encoding layers. \\
\textit{Viva la Log Fiesta} (Forensics, Beginner) & The challenge involves analyzing a log file found on a hacked server. \\
\textit{The Room} (Misc, Beginner) & A reverse-engineering/forensics challenge involving a labyrinth-like nested folder structure. \\
\textit{Bloated} (Pwn, Beginner) & A C program for counting calories; reaching 1337 calories reveals the flag. \\
\bottomrule
\end{tabularx}
\label{tab:cyberleague_challenges}
\end{table}

\begin{table}[h]
\centering
\caption{ACSC CTF Challenges employed in Ph.3 to target RQ1}
\begin{tabularx}{0.75\textwidth}{p{3cm}X}
\toprule
\textbf{CTF Challenge} & \textbf{Description} \\
\midrule
\textit{CorpoFeed} (Pwn, Beginner) & Feedback service, non-invoked win-function, ret2win \\
\textit{Cyber-Gateway} (Web, Easy) & Flask-app, restricted body contents for POST-requests \\
\textit{Nokia} (Misc, Easy) & SMS that contains three hexadecimal payloads starting with //SCKL \\
\textit{KDF\_Dream} (Crypto, Medium) & Secure messenger program with different encryption and decryption methods (DH, KDF, PRF, OTP) \\
\textit{60-seconds-to-flag} (Rev, Hard) & Windows .exe that opens a terminal and starts a countdown \\
\bottomrule
\end{tabularx}
\label{tab:ctf_challenges}
\end{table}

\subsection{Measures and Data Analysis}
\label{sec:measures}
The following measures were utilized to assess the effectiveness of solving CTF challenges with and without the use of the CAI framework. Notably, the framework was employed by a researcher with no prior penetration testing or hacking experience. In this context, the category \textit{CA(thr)I(n)} refers specifically to the researcher equipped with the CAI framework.

\subsubsection{Quantitative Measures (RQ1)}
The comparative quantitative analysis focused on the following metrics:
\begin{itemize}
    \item \textbf{Challenge Attempted:} Did they attempt the CTF challenge, and was it solved successfully? Cf. Table \ref{tab:results_yn}.
    \item \textbf{Time to Solution:} Recorded as self-reports provided by ACSC candidates. See Section \ref{sec:time-to-solution}.
    \item \textbf{Attempts to Solution:} Defined as planning and executing strategies that did not lead to success, including activities such as developing and running exploits, submitting potential flags, or testing hypotheses regarding the challenge (Section \ref{sec:strat-beh}).
    \item \textbf{Relative Time per Strategy:} The time spent on each strategy was calculated relative to the total time taken. See Section \ref{sec:strat-beh}.
\end{itemize}
\subsubsection{Qualitative Measures (RQ2)}
The qualitative analysis included the following components:

\begin{itemize}
    \item \textbf{Action Research Logs:} Records capturing the researcher's experience and processes during each challenge, including tool logs as well as field notes.
    \item \textbf{Structured Retrospective Reflection Questionnaire:} A written assessment that allowed participants to reflect on their experiences and learning, incorporating the restropective from the logs. 
    \item \textbf{Restrospective Analysis:} Data coding was done starting with an initial codebook, which was iteratively refined to identify themes related to AI delegation, output evaluation, prompt adaptation, confidence development, and ethical considerations.
\end{itemize}
For a comprehensive review of the full research instruments and coding/analysis approach, we refer interested readers to Appendix \ref{sec:appendix3}.

\subsubsection{Data Analysis}
\textbf{Quantitative} data collected from challenge attempts, solution times, and strategy execution were analyzed using descriptive statistics to compare the performance of the primary subject (equipped with CAI) to the baseline performance of last year's ACSC challengers. We report the statistical positioning of the primary subject relative to the ACSC group.
Running statistical tests, such as two-sample t-tests\footnote{The same argument holds for other tests of centrality, e.g. non-parametric tests.} or t-test testing for a fixed $\mu$ would artificially assume the individual is either a sample or a mean. It is not, it is a single case and neither represents a parameter, nor a distribution. While modified t-tests comparing simple samples to a distribution exist, given the few observations it would feel like forcing significance over a single-case focal participant study -- which is clearly not aligned with the nature of the study. Thus, we believe reporting the statistical positioning of the primary subject relative to the ACSC group yields the most methodologically adequate choice. The statistical analysis was conducted jointly by the primary subject and the second author. 

\textbf{Qualitative} data analysis, utilizing hybrid codebook/thematic analysis approach, focused primarily on the structured retrospective reflection questionnaire. The primary subject engaged in an autoethnographic process of initial documentation of their learning process, challenges encountered, and the role of AI tools in facilitating entry into penetration testing through the creation of action research logs. This documentation also provided an opportunity for critical reflection on the learning journey, and as a basis for the participant's structured self-reflection. This reflexive process enriched the data interpretation, blending structured responses with personal insights. The structured retrospective reflection questionnaire was then coded and analyzed by the second author, see \ref{sec:appendix3-codebook}.

\paragraph{(Hybrid) Reflexive Thematic Analysis} Given the autoethnographic nature of our study and its focus on personal reflection, we determined that rigidly structured, positivist coding methods were not suitable. While codebook approaches, such as Template Analysis \cite{king2017template} offer a balance between structure and qualitative research, we ultimately selected \textit{Reflexive Thematic Analysis} \cite{braun2019answers, braun2013successful} due to its prioritization of researcher reflexivity and the evolving, subjective nature of theme development. This approach best aligns with the spirit of autoethnography, acknowledging both researchers' role in shaping findings. 

Our hybrid form of reflective thematic followed Braun and Clarke's six-phase process \cite{byrne2022worked}, but with phase one skipped, directly starting with creating initial codebook (see \ref{sec:appendix3-codebook}) informed by the research question and the overarching theme of (C)AI fluency. Since the researcher coding the reflexion not only designed but accompanied the whole study, this seemed to be a natural step, which is made explicit here to adress initial relational concerns and  make the analysers familiarity with the topic and subject explicit. Moreover, their engagement with the analysis is not a limitation but a strength that adds depth to the findings. Following data familiarization, the codebook was revised, and themes were conceptualized through active researcher engagement with the data. 
\subsection{Ethics \& Limitations}
\paragraph{Ethics}
The primary study subject conducted the action research and questionnaire survey as part of a bachelor's thesis project \cite{schachner2024mastering} under the supervision of the other author. The subsequent publication occurred after the thesis submission, graduation, and exmatriculation to safeguard the primary subject and eliminate any power dependencies, ensuring they maintained full control over the publication as one of the first authors.

The anonymous, voluntary survey questionnaire (see Appendix \ref{sec:appendix3}) did not collect sensitive demographic or personal information, such as age, formal education, or gender. According to institutional policy, a formal ethics board review was not required. Nevertheless, the survey conducted among ACSC participants adhered to the same ethical and GDPR standards as studies requiring formal review, ensuring participant anonymity throughout recruitment, data collection, and analysis. The questionnaire link was shared via the ACSC Discord channel and included a consent form stating that participation was voluntary. It specified the data's use for this study and confirmed the confidentiality and GDPR compliance of the responses, including rights such as deletion and information of use and secure storage. Participants were free to withdraw from the study at any time. The survey concluded with a debriefing statement and contact information for the corresponding author. The questionnaire was securely hosted on our institution's LimeSurvey server to maintain the integrity and confidentiality of the collected data. For reproducibility, the processed anonymous dataset is attached. 

\paragraph{Use of Generative AI}
In addition to employing Generative AI (GenAI) as an action research tool under study (cf. Methodology Section), LLMs such as GPT-4o mini, ChatGPT 5.2, and Gemma 3-27B were utilized for editorial purposes. Importantly, GenAI was not used for literature research, study design, methodological choices, or evaluations. Before submitting the manuscript, GenAI served as an advocatus diaboli to critically examine the methodology and evaluation processes for limitations. This approach enabled the authors to reflect on strengths and weaknesses, leading to a deeper understanding of the research implications. All outputs were carefully reviewed to ensure accuracy and originality.

\paragraph{Methodological Limitations} Since the study utilized an anonymous and voluntary survey, the sample of participants who chose to respond may not be fully representative of the broader population. This self-selection bias can lead to several limitations, including demographic variation \footnote{Individuals with a specific interest or confidence in cybersecurity and AI tools might have been more likely to participate, potentially skewing the results.} or social desirability bias. \footnote{Hidden in an anonymous crowd, the survey participants might give socially desirable answers or overstate their actual level of expertise.} Moreover, we highlight the inherent limits of the autoethnography inspired action-research methodology that is part of our research design. 

This study's initial focus centered on a single participant – a student with a systems engineering background, but lacking prior experience in Capture The Flag (CTF) competitions. This individual provided valuable insights into the process of learning to utilize cybersecurity AI tools, especially for especially for skilled users who transition form other fields in IT to cybersecurity. The pre-existing technical foundation and problem-solving skills may, however, not be representative of the broader population.  
Specifically, their rapid adaptation to the tool could be attributed to transferrable skills rather than a general ease of use for all users. This presents a limitation in generalizing the findings to individuals without similar technical backgrounds – for example, those without formal engineering training or those new to both cybersecurity  and systems-level thinking. 

The study was conducted using a specific agentic AI system, specifically CAI mostly combination with Anthropic Models Claude Haiku 3.5 \cite{anthropic2024haiku} and Sonnet 3.7 \cite{anthropic2025sonnet} with a defined setup. The observed performance and user experience are tied to this particular implementation, and may not generalize to other LLMs or agentic AI toolchains. Future research should explore the impact of using different LLMs and tool combinations to determine the broader applicability of these findings.

By acknowledging these limitations of our methodology, we recognize the constraints of our findings and suggest further research should include different cybersecurity AI framworks as well as participants with diverse backgrounds and skills to research generalizability. This includes individuals with limited prior technical experience, as well as those with experience in other fields beyond systems engineering, to enhance the validity and applicability of the results.

\section{Results}
\subsection{Relative Performance Measures}
This section presents the comparative benchmarking adressing RQ1-- in terms of time, attempts and time per attempt required for each task -- of solving the discussed CTF challenges. Recall that the primary test subject possessed no prior penetration testing or hacking experience, but has a systems engineer background and was equipped but not familiar \textit{with} the CAI framework. The label \textit{CA(I)thrin}\footnote{CA(I)thrin is a mixture of CAI \cite{mayoralvilches2025caiopenbugbountyready} and the working of the study \textbf{C}an \textbf{AI} lower \textbf{TH}e bar\textbf{RI}er to cybersecurity for \textbf{N}ovices? -- CA(I)thrin.} is an acronym and refers to the researcher and primary subject equipped with the CAI framework. The aim is to report the positioning of the primary subject relative to the ACSC CTF Challengers.

\begin{table}[h]
\centering
\caption{Solve rates ($N=$ rel. sample size)}
\begin{tabularx}{0.75\textwidth}{p{1.6 cm}XXXXX}
\toprule
\textbf{Group} &  \rotatebox{45}{\textbf{CorpoFeed}}&\rotatebox{45}{\textbf{Cyber-Gateway}}&\rotatebox{45}{\textbf{Nokia}}&\rotatebox{45}{\textbf{KDF\_Dream}}&\rotatebox{45}{\textbf{60-sec...}}\\
\midrule
\textit{All Levels} & 89\% (18)& 100\% (7)& 100\% (10)& 25\% (4) & 50\% (4)\\
\midrule
\textit{Beginner} & 50\% (4)& - & 100\% (2)& - & -\\
\textit{Intermediate} & 100\% (8)& 100\% (6)& 100\% (7) & 33\% (3)& 50\% (4)\\
\textit{Advanced} & 100\% (3)& - & - & - &- \\
\textit{Expert} & 100\% (3)& 100\% (1)& 100\% (1)& 100\% (1)& -\\
\midrule
\textit{CA(I)thrin}  & $\checkmark$ & $\checkmark$ & $\checkmark$ & $\times$ & $\checkmark$ \\
\bottomrule
\end{tabularx}
\label{tab:results_yn}
\end{table}

For readability, the quantitative measures, where the primary subject performed better or equal than the mean, are marked with a dagger $\dagger$ in tables \ref{tab:results_time},\ref{tab:results_strt},\ref{tab:results_TA}. An overview of the overal solve rates is given in Table \ref{tab:results_yn}.

\subsubsection{Time to Solution}
\label{sec:time-to-solution}
Overall, the primary subject demonstrated average efficiency in the \textit{Cyber-Gateway} and \textit{Nokia} CTFs, but faced significant challenges in the \textit{CorpoFeed} and \textit{KDF\_Dream} tasks, indicating a mixed performance profile with respect to overall time to solution, influenced by the complexity of each challenge. Specifically, in \textit{CorpoFeed}, the primary subject had to perform certain tasks manually (calculating offset), as the framework was unable to utilize GDB correctly. This additional manual intervention contributed to increased time and number of attempts recorded for \textit{CA(I)thrin} in Table \ref{tab:results_time}.
\begin{table}[h]
\centering
\caption{Mean Time to Solution [in h] ($\pm$ std. deviation))}
\begin{tabularx}{0.75\textwidth}{p{1.6cm}XXXXX}
\toprule
\textbf{Group} &  \rotatebox{45}{\textbf{CorpoFeed}}&\rotatebox{45}{\textbf{Cyber-Gateway}}&\rotatebox{45}{\textbf{Nokia}}&\rotatebox{45}{\textbf{KDF\_Dream}}&\rotatebox{45}{\textbf{60-sec...}}\\
\midrule
\textit{All Levels} & 1.78 ($\pm$1.76)& 0.83 ($\pm$1.07)& 2.29 ($\pm$3.1)& 3.25 ($\pm$1.71)&7.67 ($\pm9.07$)\\
\midrule
\textit{Beginner}  &3.0 ($\pm 1.73$)& -& 5.12 ($\pm$6.89)&- &-\\
\textit{Intermediate} &1.69 ($\pm1.98$) & 0.95 ($\pm1.08$)& 1.67 ($\pm1.75$)& 4.0 ($\pm1$)&7.67 ($\pm9.07$)\\
\textit{Advanced} & 1.0 ~~ ~~(--)& -& -& -&-\\
\textit{Expert} & 0.62 ~~($\pm$0.53) & 0.1 ~~ ~~(--)& 1.0 ~~ ~~(--)& 1.0 ~~ ~~ (--)&-\\
\midrule
\textit{CA(I)thrin} & 7.0 &1.0 &\textbf{1.0}$^\dagger$&10.0 &\textbf{4.0}$^\dagger$\\
\textit{\%-rank} & $>$100\% & 75\% & 44\%&$>$100\% & 50\%\\
\bottomrule
\end{tabularx}
\label{tab:results_time}
\end{table}

\subsubsection{Strategic Behavior}
\label{sec:strat-beh}
\paragraph{Strategy Exploration}In relative performance, the primary subject consistently demonstrated a tendency to explore more strategies, resulting in a higher number of attempts compared to ACSC Challengers and placing the primara subject after or equal to the 75th percentile in all cases, see table \ref{tab:results_strt}. The \textit{60-Seconds-To-Flag} challenge saw the primary subject needing roughly ten attempts versus five for intermediates. The \textit{KDF\_Dream} challenge was particularly revealing, as it remained unsolved even after twenty attempts and ten hours, starkly contrasting with experts who took only one attempt. Nontheless, from a didactic point of view, learners could benefit from the exploration behavior on a strategic level, exploring various strategies at a fast rate.
\paragraph{Per-Strategy Timing}
In most challenges, the primary subject achieved a Time per Attempt (T/A) better than average, even  outperforming the averages of beginners and intermediates, still lagging behind experts in many cases. 

Our main observation is, that comparatively quick per-attempt times were achieved across challenges, while usually higher than all number of strategies were explored. This observation implies that AI agentic tools could facilitate learning on a rather strategic and tactical level rather than (purely) operational. Individuals with adjacent technical backgrounds but limited exposure to offensive security practice might therefore particulary benefit from agentic- cybersecurity framework use, e.g. for security training or CTF entry.
\begin{table}[h!]
\centering
\caption{Avg. Number of Attempts ($\pm$ std. deviation)}
\begin{tabularx}{0.75\textwidth}{p{1.6cm}XXXXX}
\toprule
\textbf{Group} &  \rotatebox{45}{\textbf{CorpoFeed}}&\rotatebox{45}{\textbf{Cyber-Gateway}}&\rotatebox{45}{\textbf{Nokia}}&\rotatebox{45}{\textbf{KDF\_Dream}}&\rotatebox{45}{\textbf{60-sec...}}\\
\midrule
\textit{All Levels} & 4.38 ($\pm 5.01$)& 1.71 ($\pm 0.95$)& 2.66 ($\pm2.06$)& 2.25 ($\pm1.5$)& 5.0 ~ ($\pm3.91$)\\
\midrule
\textit{Beginner}  &7.66 ($\pm$6.66) & -& 5.0 ~~ ~~(--)&- &-\\
\textit{Intermediate} &3.71 ($\pm$5.02) & 1.83 ($\pm 0.98$)& 2.57 ($\pm 1.92$)& 2.66 ($\pm 1.25$)& 5.0 ($\pm 3.92$)\\
\textit{Advanced} & 1.0 ~~ ~~ (--)& -& -& -&-\\
\textit{Expert} & 3.5 ~~($\pm$3.54) & 1.0 ~~ ~~(--)& 1.0 ~~ ~~(--)& 1.0 ~~ ~~ (--)& -\\
\midrule
\textit{CA(I)thrin} & 19.0 &5.0 &5.0 &20.0 &10.0 \\
\textit{\%-rank} & $>$100\% & $>$100\% & 75\% &$>$100\% & $100\%$\\
\bottomrule
\end{tabularx}
\label{tab:results_strt}
\end{table}

\begin{table}[h]
\centering
\caption{Mean Time per Attempt [in h] ($\pm$ std. d.). $^\ddag$ Two extreme outliers, i.e. id 12, 13, were detected. Removing them yields a mean of $0.54$ and empirical standard deviation $s=0.28)$ for all levels and a mean of $0.61$ and empirical standard deviation $s=0.24)$ for intermediates. The primary subject did not solve this CTF.}
\begin{tabularx}{0.75\textwidth}{p{1.6cm}XXXXX}
\toprule
\textbf{Group} &  \rotatebox{45}{\textbf{CorpoFeed}}&\rotatebox{45}{\textbf{Cyber-Gateway}}&\rotatebox{45}{\textbf{Nokia}}&\rotatebox{45}{\textbf{KDF\_Dream}}&\rotatebox{45}{\textbf{60-sec...}}\\
\midrule
\textit{All Levels} & 0.91$^\ddag$ ($\pm 1.55$)& $0.37$ ($\pm 0.31$)& $0.76$ ($\pm 0.52$)& $1.65$ ($\pm 0.91$)& $1.24$ ($\pm 1.52$)\\
\midrule
\textit{Beginner}  &$0.47$ ($\pm 0.2$)& -& 2.0 ~~ ~~ (--)&- &-\\
\textit{Intermediate} &1.29$^\ddag$  ($\pm 2.1$) & 0.4 ($\pm 0.32$)& 0.55 ($\pm 0.22$)&  1.87 ($\pm 0.99 $)& 1.24 ($\pm1.52$)\\
\textit{Advanced} & 1.0 ~~ ~~(--)& -& -& -&-\\
\textit{Expert} & 0.21 ~~($\pm 0.06$) & 0.1  ~~ ~~(--)& 1.0 ~~ ~~(--)& 1.0 ~~ ~~ (--)&-\\
\midrule
\textit{CA(I)thrin} & \textbf{0.36}$^\dagger$ &\textbf{0.2}$^\dagger$ &\textbf{0.2}$^\dagger$ &\textbf{0.5}$^{\dagger \ddag}$ &\textbf{0.4}$^\dagger$  \\
\textit{\%-rank} & 32\% & 40\% & $<$0\%& $<$0\%& 50\%\\

\bottomrule
\end{tabularx}
\label{tab:results_TA}
\end{table}

\subsection{AI-Mediated Entry to Cybersecurity Practise: Thematic Analysis}
To answer RQ2, the structured restrospective reflection questionnaire answered by the primary subject was coded and thematic analysis performed. The thematic analysis reveals a progression from \textit{perceived cognitive entry barriers} to a \textit{transformation in skill}, \textit{confidence}, and \textit{professional self-conception}. Across all identified themes, the agentic AI framework under study is not just an automation tool, but a mediator of orientation, structured engagement, and reflective learning.

\subsubsection{Cognitive Entry Barriers to Hacking}
Before the study, the primary subject perceived CTF participation as inaccessible and intimidating: ``Without structured guidance, the entry threshold appeared high and discouraging.'' Two subthemes were identified: The first is \textit{Elitism Perception}, where cybersecurity practise and hacking is perceived as exclusive to highly specialized experts. The  candidates lack of prior exposure and practical experience solidified this perception, resulting in reluctance to engage in CTFs.
The second subtheme -- \textit{Skills Gap} -- subsumes missing practical and methodolocial skills. The primary subject revealed not knowing ``how and where to start to solve a CTF'', calling it ``overwhelming'' in the ``absence of methodological structure'' guidance. Thus, the missing procedural orientation constituted a significant barrier for the else technically versatile participant. 

Our themes illustrate that the primary entry barrier was not a technical one, but cognitive and structural one: doing CTFs simply appeared complex, opaque, exclusive.

\subsubsection{Agentic Cybersecurity AI as Mental Mapping and Orientation}
The introduction of the agentic AI framework CAI altered this perception. Three subthemes characterize this mediation: \textit{Strategic Overview, Strategic Guidance}, and \textit{Cognitive Load Reduction}.

First, the cybersecurity AI provides a mental map of possible approaches (\textit{Strategic Overview}). Next, instead of exploratory uncertainty, the AI system provided \textit{Strategic Guidance}, by analysing the context, identifying  potential vulnerabilities, structured, hypothesis-driven reasoning. It extendes beyond orientation, since CAI effectively structuring workflow by in sequencing tasks and suggesting appropriate tools. This distinction between \textit{Strategic Overview} and \textit{Guidance} is crucial: one facilitates cognitive mapping, the other  procedural execution.

Finally, \textit{Cognitive Load Reduction} emerged as a central effect. Especially in early stages, AI-supported structuring reduced feelings of overwhelm. Dissassembling complex problems into small steps caused decreasing uncertainty and higher motivation. Some might argue this eliminate efforts, others that this frees mental capacities for analysis rather than navigation.

Overal, our findings suggest that the AI functioned as an orientation, lowering the cognitive barrier for entry into offensive security tasks.

\subsubsection{Knowledge Gain: Procedural and Conceptual}
The reduction of cognitive load enabled substantive learning outcomes. Two complementary forms of knowledge gain emerged: \textit{Practical Skill Acquisition and Transferability}, as well as \textit{Conceptual Understanding}.

The primary subject reportedly gained practical skills in reconnaissance, vulnerability identification, and tool usa. Moreover, partial transferability was indicated: beginner-level tasks can now be performed independently, while advanced techniques potentially require assistance.

Additionally, the AI-supported explanations facilitated \textit{Conceptual Understanding}. Rather than full automation, the system made the underlying mechanisms explicit, facilitating connections between theoretical and applied knowledge. This suggesting that - for the particular tool under study - agentic AI use can foster deeper cognitive integration via structured learning through guided interaction instead of passive task completion.

\subsubsection{Identity Shift and Transformation}
In our study, this resulted in an \textit{Identity Shift}.
While offensive cybersecurity practise was deemed as a mysterious, elite domain before the study, the primary subject described it as structured, learnable, and accessible through guided exploration after the study. This shift encompassed a shift in professional self-perception, icluding an increased CTF participation readiness, as well as higher motivation and confidence.

Importantly, the transformation was not framed as mastery, but sophisticated entry into the field. The primary subject moved from perceiving hacking as inaccessible to viewing it as a domain navigable through analytical reasoning and structured practice. This suggests that AI-mediated orientation may not only influence performance, and provide faster and enhanced strategy exploration (cf. \ref{sec:measures}), but also reshape professional identity formation in offensive security contexts.
\subsubsection{AI Ownership vs. Automation Risk}
Alongside these positive developments, the analysis revealed an ongoing tension between AI Ownership and Automation Risk. The primary subject emphasized the necessity of critical evaluation of AI outputs, particularly when suggestions were generic or misaligned with the task context. Prompt refinement or independent investigation indicate emerging \textit{AI fluency competencies}\cite{mayoralvilches2025caifluencyframeworkcybersecurity}, such as discernment and strategic delegation. Still, the moments of increased reliance were also acknowledged, particularly when encountering repeated obstacles. The candidate also adresses the danger of mental offloading and limited learning effects as well as concerns, if AI suggestions were followed without discernment. This reflection highlighted concerns about responsible use, particularly in real-world contexts requiring authorization and legal boundaries.

Thus, AI mediation did not eliminate critical reasoning, but it introduced a new domain of responsibility: managing the balance between assistance and autonomy.

\section{Conclusion}
This work addresses the question whether agentic systems function merely as automation tools - or as facilitators of early-stage learning. Studying the cybersecurity AI framework CAI \cite{mayoralvilches2025caiopenbugbountyready}, we seeked to accompany a novice's entry into Capture-the-Flag (CTF) practice in a longitudinal human-centered, mixed-methods case study. Rather than benchmarking agentic cybersecurity frameworks performance in isolation, we positioned an AI-assisted novice within a national CTF challenges performance distributions (without agentic AI) and analyzed the experiential dimensions of AI-supported CTF Learning and performance.

Quantitatively, the primary subject's performance was situated within the range of intermediate competitors across several challenges, while exhibiting extensive strategic exploration and comparatively low per-strategy execution times. These findings suggest that AI assistance may enable early engagement at a strategic level, even in the absence of prior CTF experience.

Qualitatively, the analysis revealed a processual transformation of the candidate: while perceived cognitive barriers and elitism toward hindered CTF entry before the study, leveraging the CAI tool not only helped gain procedural and conceptual knowledge, but increase confidence. Rather than replacing learning, agentic cybersecurity AI frameworks, such as CAI, have the potential to restructure early-stage CTF participation by providing structure and overview, preventing cognitive overload of novices. In our observations, the framework functioned less as an automation substitute but as an orientation and structure guide, providing a mental mapping, workflow structuring, thus reducing cognitive friction in early engagement. At the same time, tensions related to overreliance, trust calibration as well as responsible use highlight the need for careful integration of such systems into educational contexts.

Taken together, our findings suggest that agentic cybersecurity AI frameworks may reshape early-stage participation in offensive security by redistributing cognitive demands and compressing aspects of the apprenticeship phase. While limited to a single-case design of a novice-but-skilled IT professional, this study contributes empirical grounding to ongoing discussions about AI-mediated cybersecurity training, and identifies key directions for future research on sustainable and responsible integration.


\section*{Acknowledgments}
We gratefully acknowledge the support of Cybersecurity Austria, and specifically Manuel Reinsperger, for the valuable feedback and assistance with the survey distribution through ACSC channels. We also extend our sincere thanks to Aliasrobotics, and particularly V. M. Vilches and C. Veas, for their invaluable support and early access to CAI and comprehensive documentation. This non-monetary cooperation facilitated the prompt initiation of this study and ultimately led to the development of an online series dedicated to learning CTF challenges with agentic AI, and the publication of the technical report CAI Fluency accompanying the tutorials.

\bibliographystyle{plain}
\bibliography{references}




\appendix
\onecolumn

\section{Appendix 1: Related Concepts and Definitions}
\label{sec:appendix1}
\subsection{CTF Categories}
This section outlines the principal categories used in CTF competitions and adopted throughout this research project. 
\begin{table}[h]
    \centering
    \caption{CTF Challenge Categories}
    \begin{tabularx}{0.99\textwidth}{p{4cm}X }
        \toprule
        \textbf{CTF Challenge Category} & \textbf{Description} \\
        \midrule

        \textit{Rev - Reverse Engineering} & 
        \textit{Reverse engineering} tasks involve analyzing compiled programs to recover program logic and derive inputs or flags, typically using disassemblers and decompilers \cite{CTF2024Rev}. \\

        \textit{Web} & 
        \textit{Web} challenges target vulnerabilities in web applications and APIs (e.g., injection, authentication flaws) requiring exploitation to access data or execute code. Techniques align with the OWASP Top 10 \cite{Owasp2025Web}. \\

        \textit{Crypto - Cryptography} & 
        \textit{Cryptography} themed CTFs practice applied cryptography. Challenges demand using algorithms and theoretical concepts to recover secrets or break flawed constructions \cite{Nagare2025Crypto}. \\

        \textit{Pwn - Binary Exploitation} & 
        \textit{Binary exploitation} exercises involve finding and exploiting bugs in binaries (e.g., overflows) to gain code execution or read protected data \cite{CTF2024Pwn}. \\

        \textit{Forensic} & 
        \textit{Digital forensics} is the systematic analysis of digital artifacts to reconstruct past activity. Tasks involve disk images, logs, and metadata, with flags often concealed in non-obvious locations \cite{CTF2024Forensics}. \\

        \textit{Misc - Miscellaneous} & 
        \textit{``Misc''} includes problems that do not fit the other categories, ranging from scripting to steganography and data recovery. Misc tests breadth, creativity, and tooling fluency \cite{Springer2022Misc}. \\
        \bottomrule
        
    \end{tabularx}
    \label{tab:acsc_ctf_challenges}
\end{table}
\newpage
\section{Appendix 2: Supplementary Data}
\label{sec:appendix2}
In the following Sections, histograms of the quantitative measures for visual comparison are provided. 
\begin{figure}[h!]\begin{center}
        \includegraphics[width=0.88\textwidth]{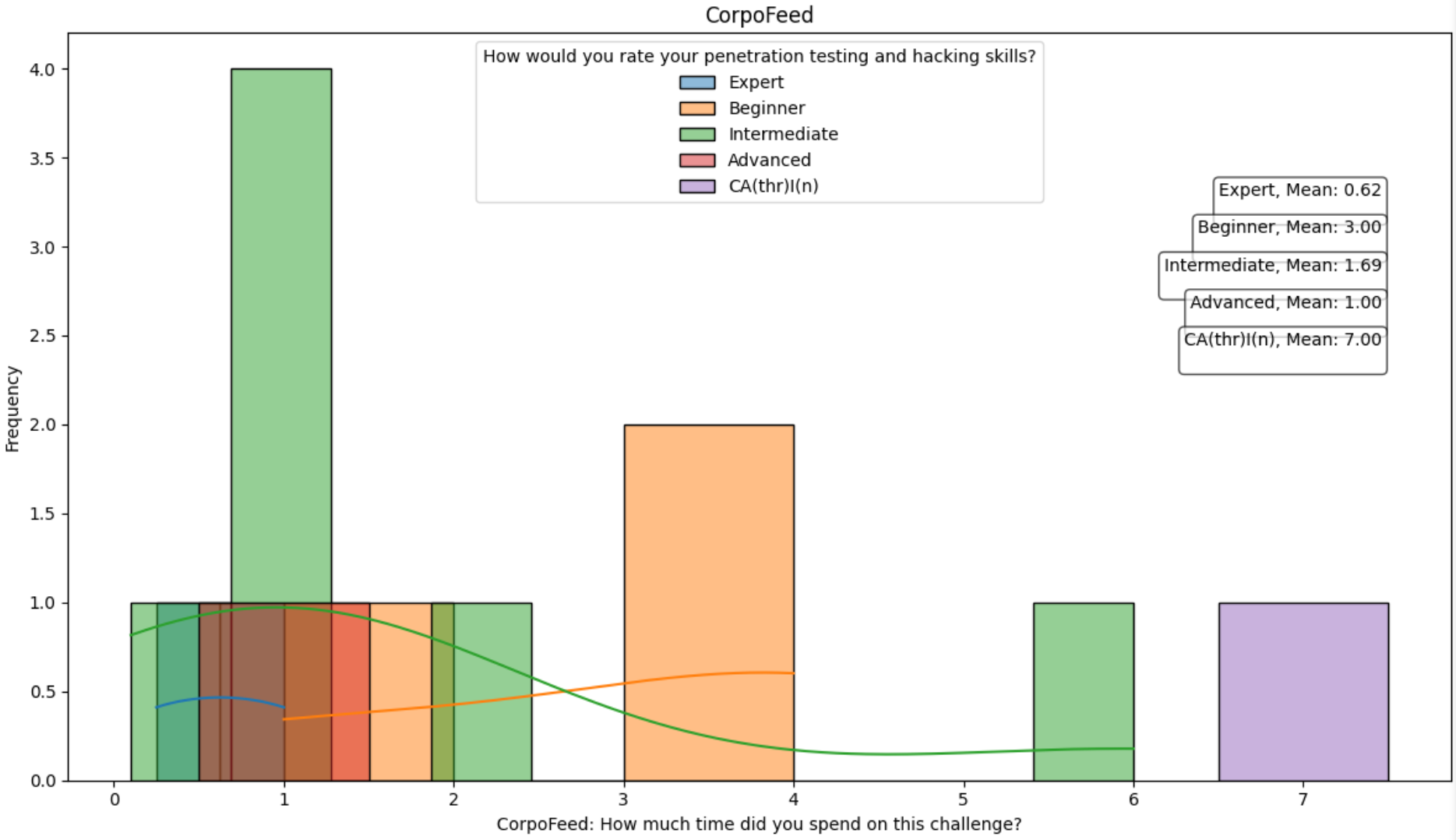}
        \caption{CorpoFeed: Time-to-completion --Histogram}
        \label{fig:cai_corpo_time}
      \end{center} \end{figure}
      
 \begin{figure}[h!]\begin{center}
        \includegraphics[width=0.88\textwidth]{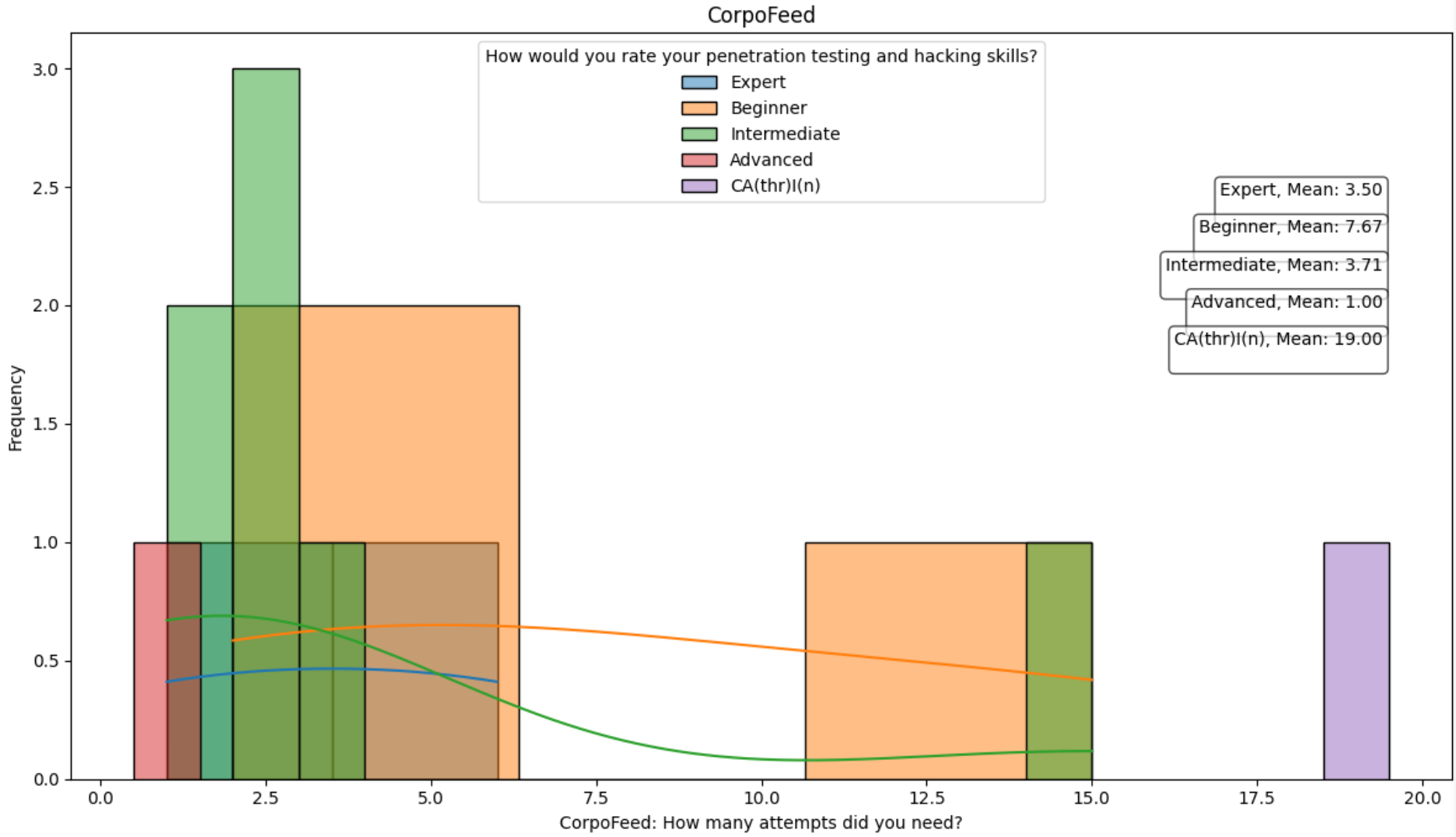}
        \caption{CorpoFeed: Number of Attempts needed -- Histogram}
        \label{fig:cai_corpo_attempts}
      \end{center} \end{figure}

\renewcommand{\arraystretch}{1.5}

 \begin{figure}[h!]\begin{center}
        \includegraphics[width=0.88\textwidth]{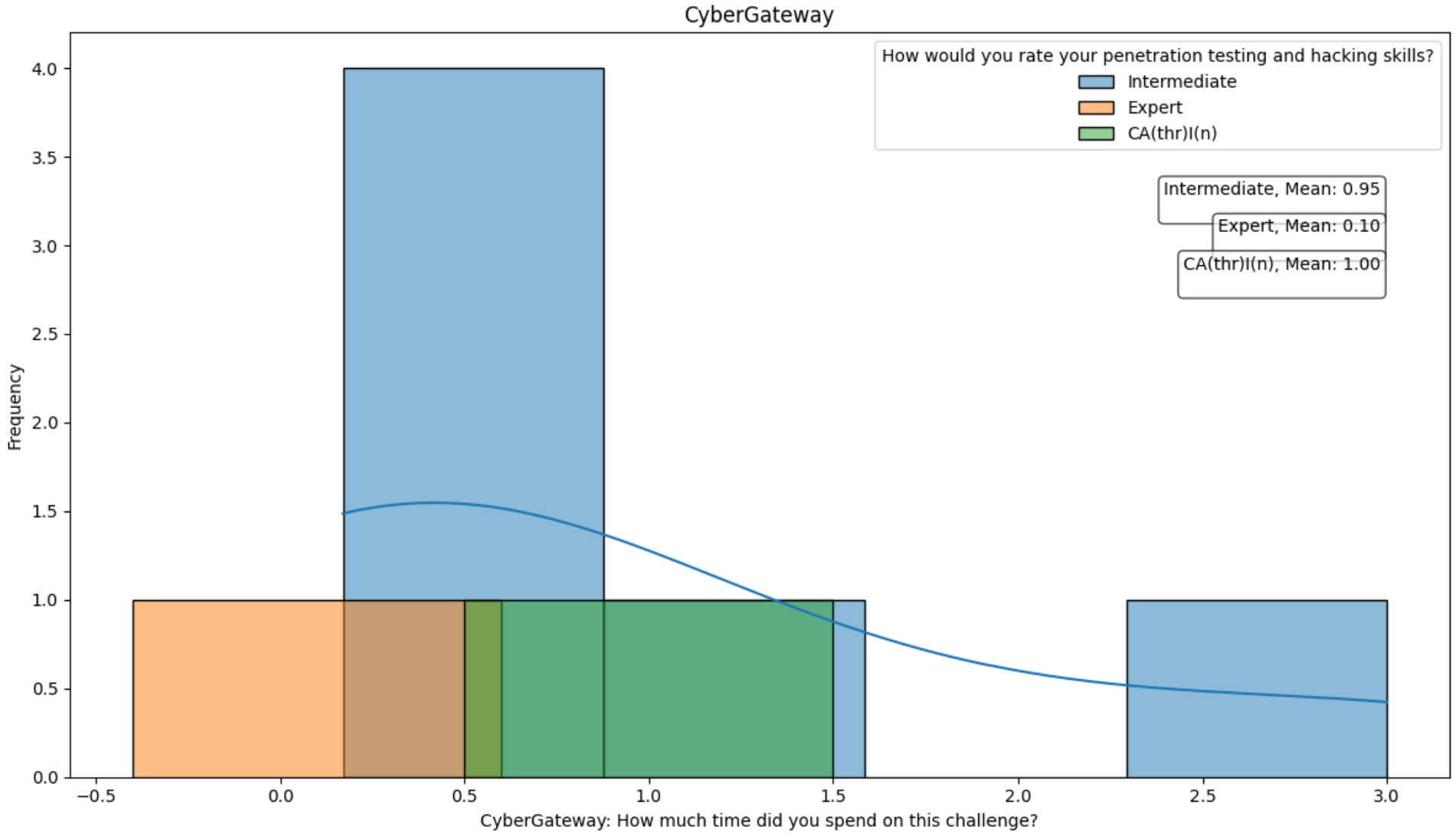}
        \caption{Cyber-Gateway: Time-to-completion --Histogram}
        \label{fig:cai_cyber_time}
      \end{center} \end{figure}
      
 \begin{figure}[h!]\begin{center}
        \includegraphics[width=0.88\textwidth]{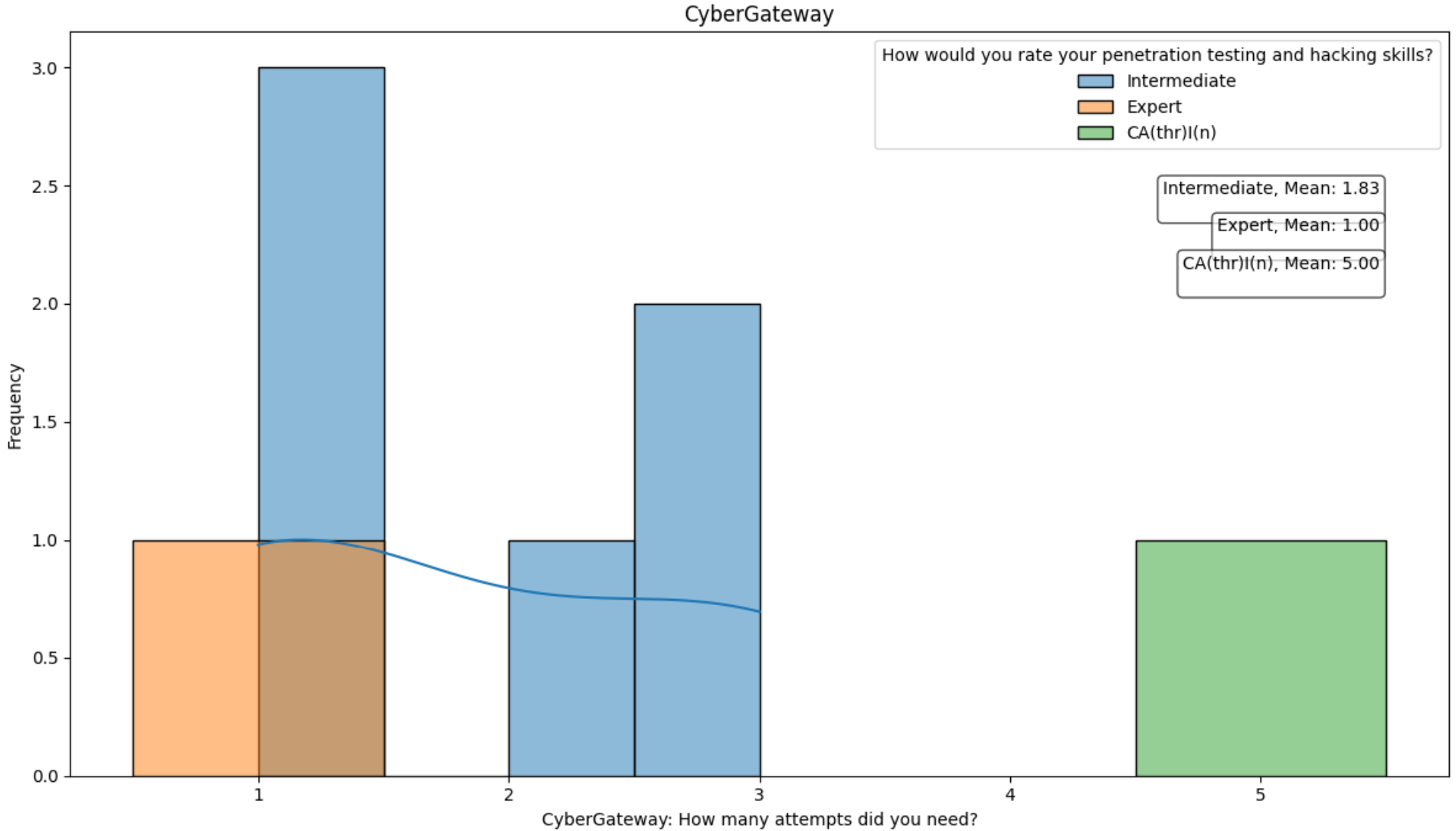}
        \caption{Cyber-Gateway: Number of Attempts needed -- Histogram}
        \label{fig:cai_cyber_attempts}
      \end{center} \end{figure}

\renewcommand{\arraystretch}{1.5}
 \begin{figure}[h!]\begin{center}
        \includegraphics[width=0.88\textwidth]{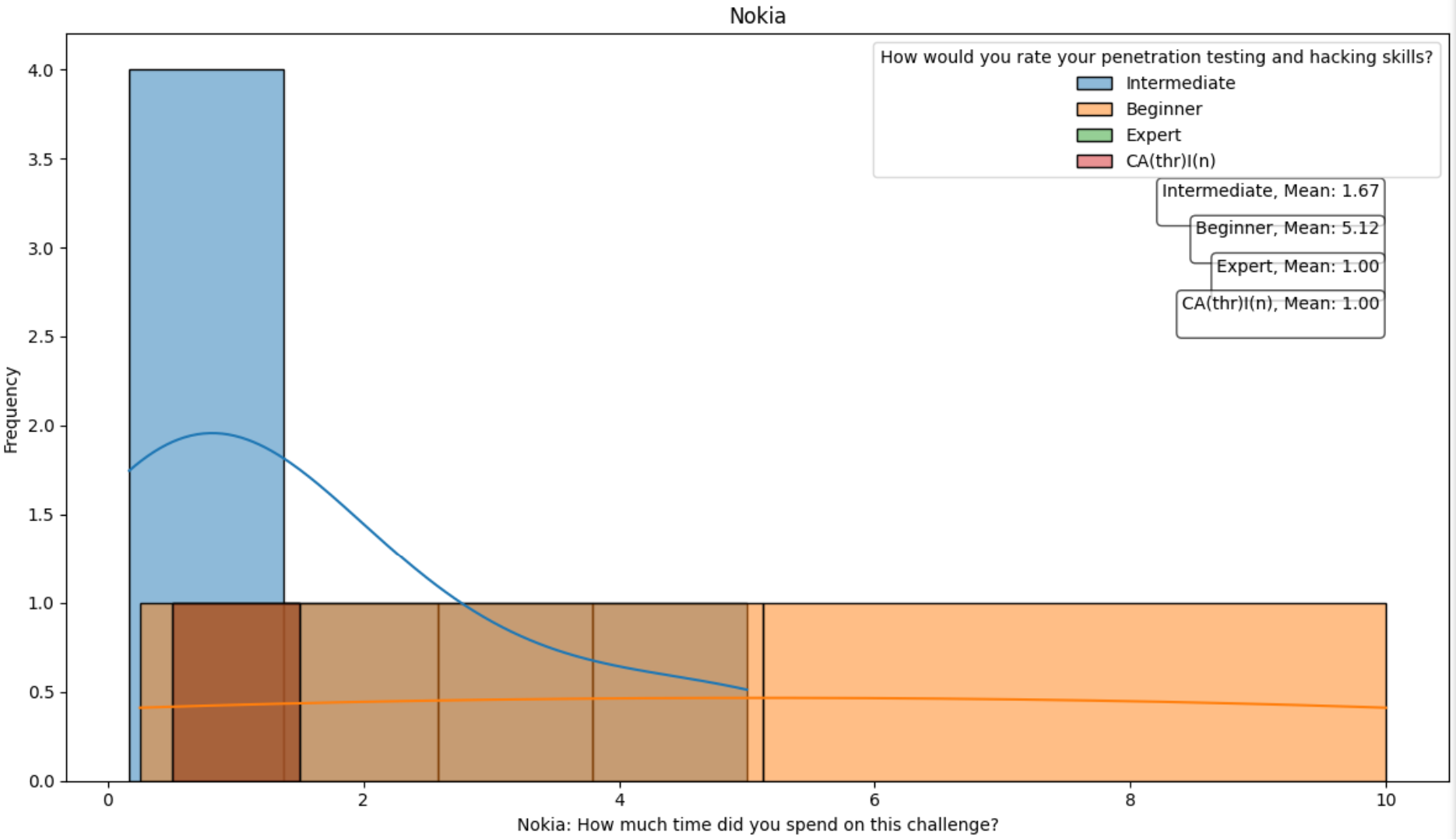}
        \caption{Nokia: Time-to-completion --Histogram}
        \label{fig:cai_nokia_time}
      \end{center} \end{figure}
      
 \begin{figure}[h!]\begin{center}
        \includegraphics[width=0.88\textwidth]{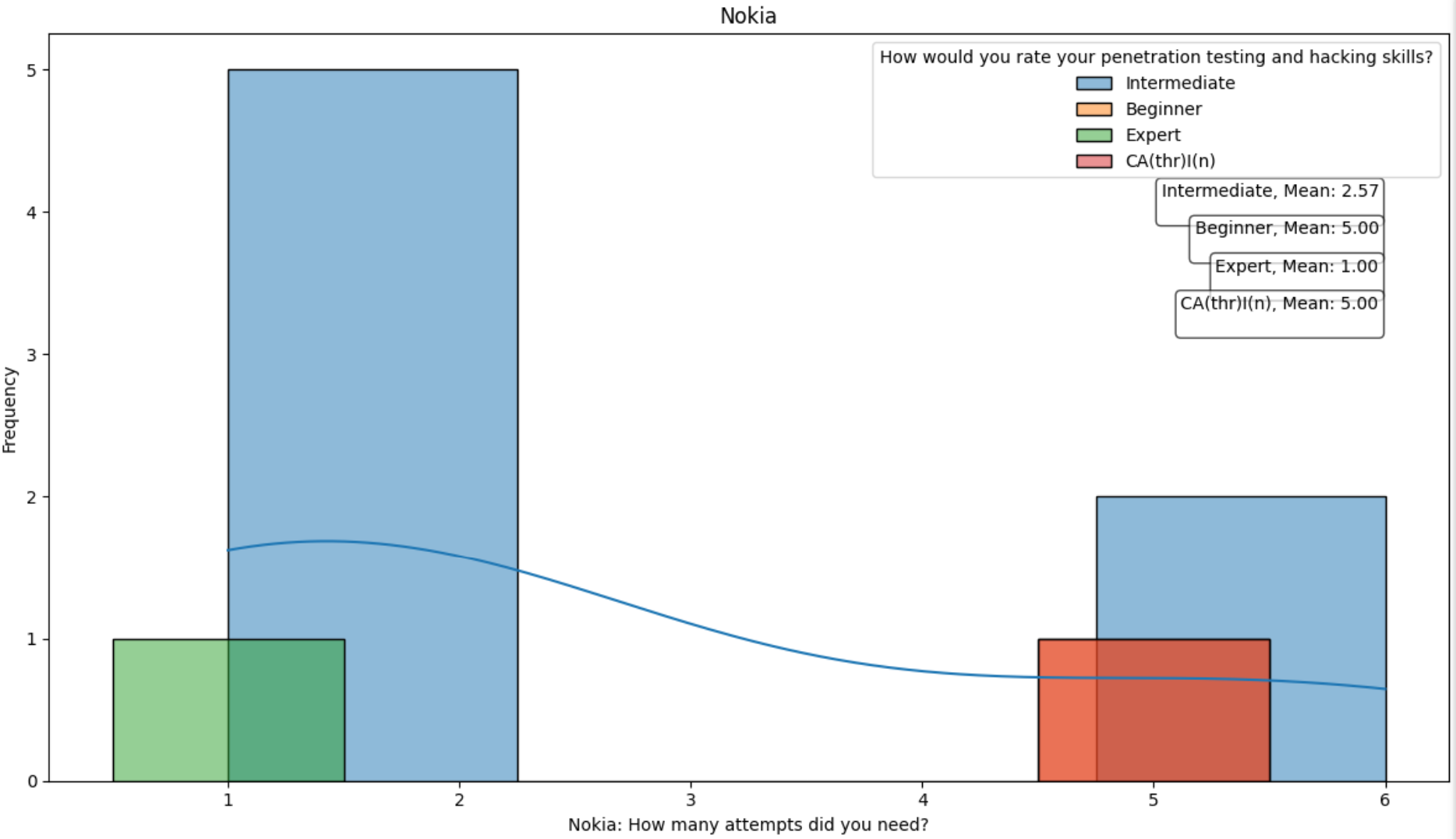}
        \caption{Nokia: Number of Attempts needed -- Histogram}
        \label{fig:cai_nokia_attempts}
      \end{center} \end{figure}

\renewcommand{\arraystretch}{1.5}
 \begin{figure}[h!]\begin{center}
        \includegraphics[width=0.88\textwidth]{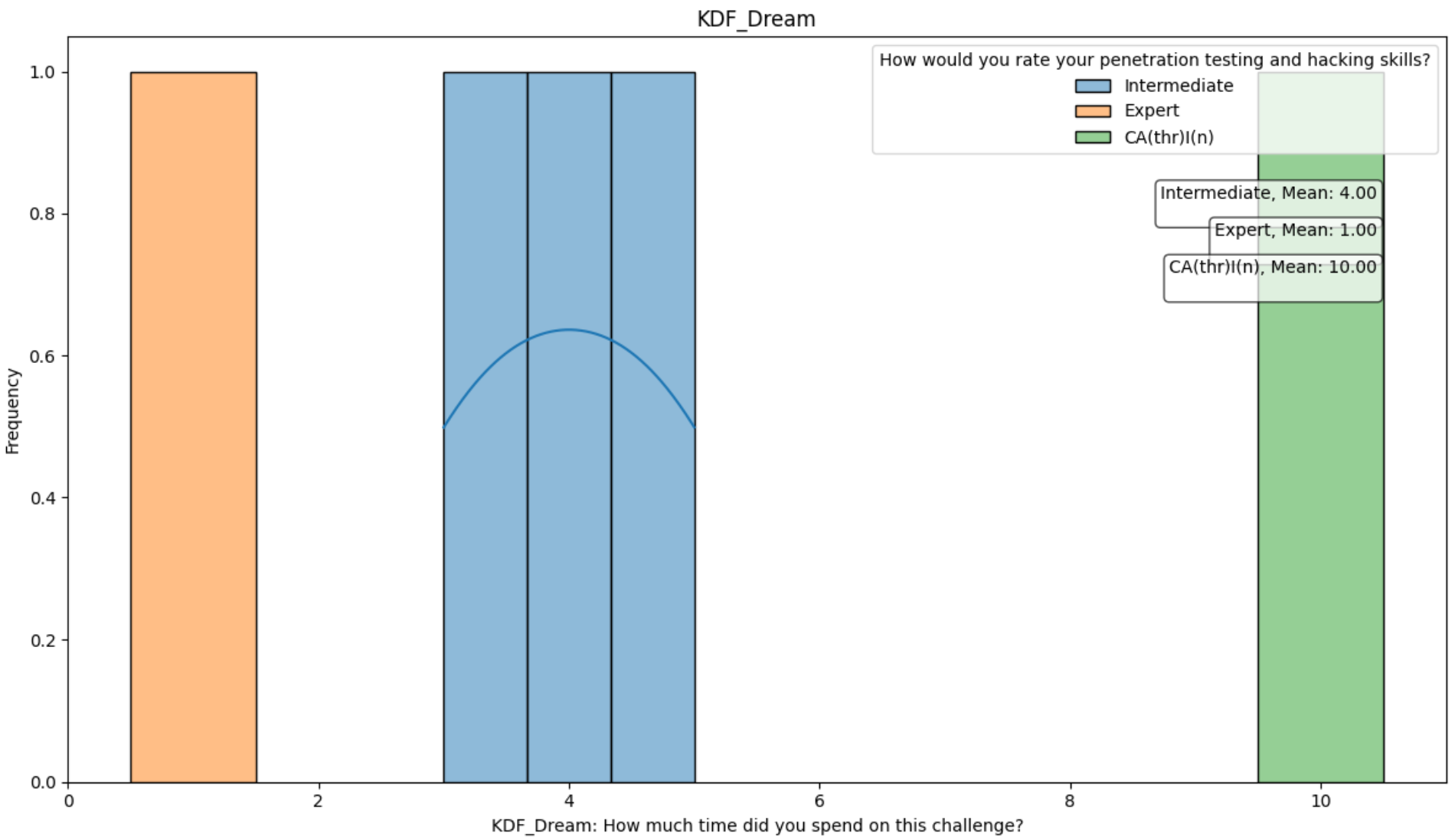}
        \caption{KDF\_Dream: Time-to-completion --Histogram}
        \label{fig:cai_kdf_time}
      \end{center} \end{figure}
      
 \begin{figure}[h!]\begin{center}
        \includegraphics[width=0.88\textwidth]{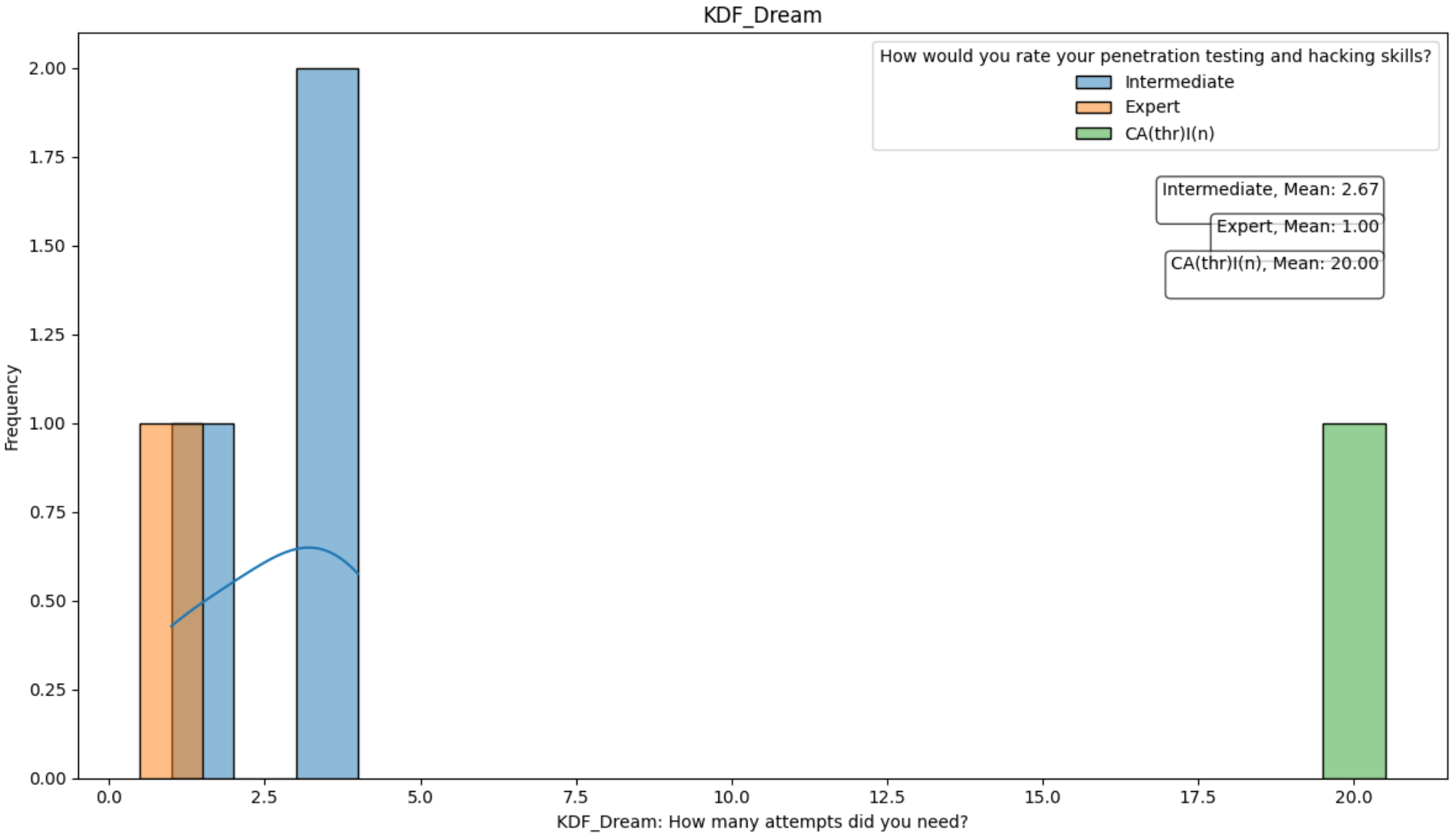}
        \caption{KDF\_Dream: Number of Attempts needed -- Histogram}
        \label{fig:cai_kdf_attempts}
      \end{center} \end{figure}


 \begin{figure}[h!]\begin{center}
        \includegraphics[width=0.88\textwidth]{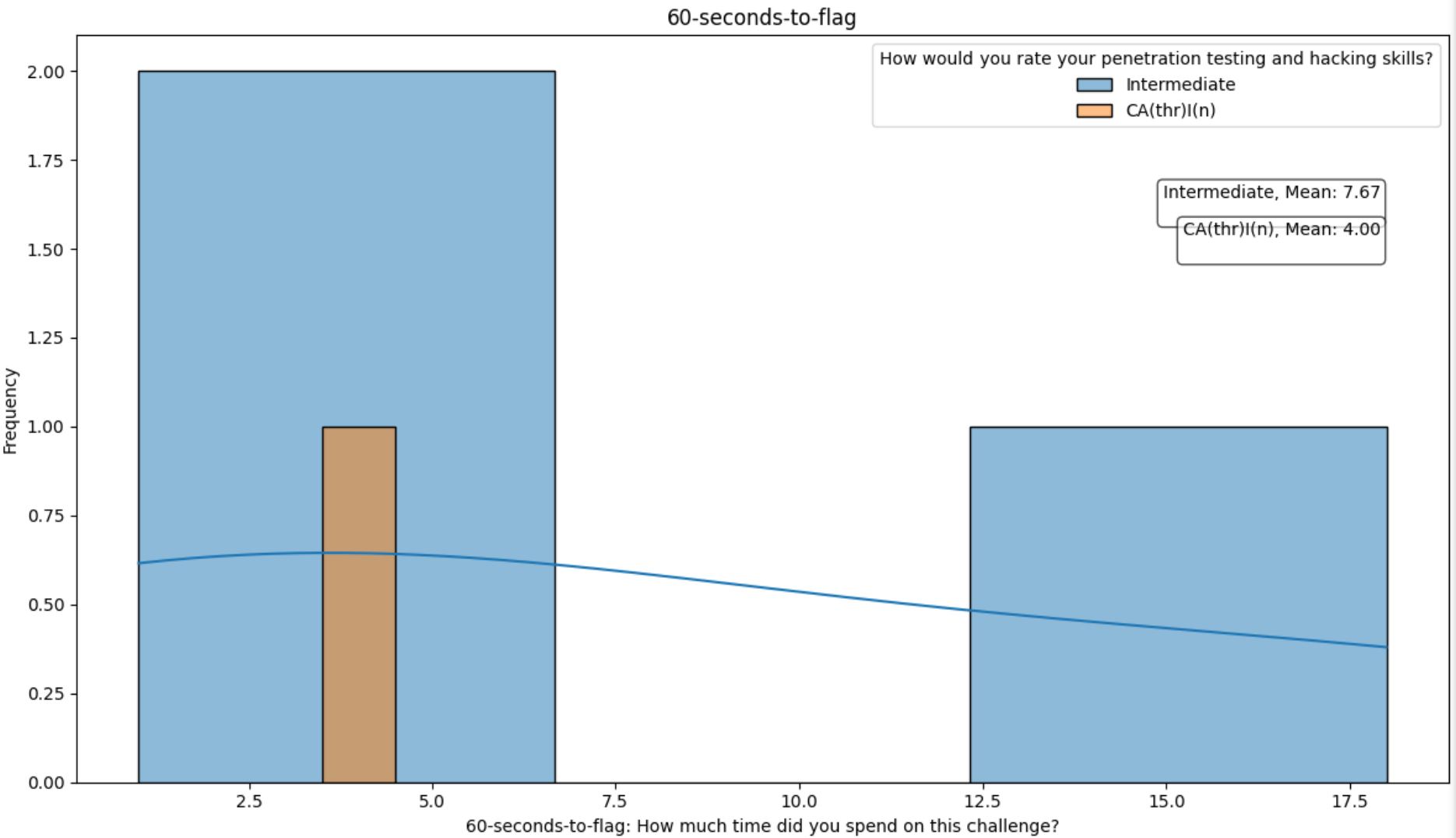}
        \caption{60-seconds-to-flag: Time-to-completion --Histogram}
        \label{fig:cai_60_time}
      \end{center} \end{figure}
      
 \begin{figure}[h!]\begin{center}
        \includegraphics[width=0.88\textwidth]{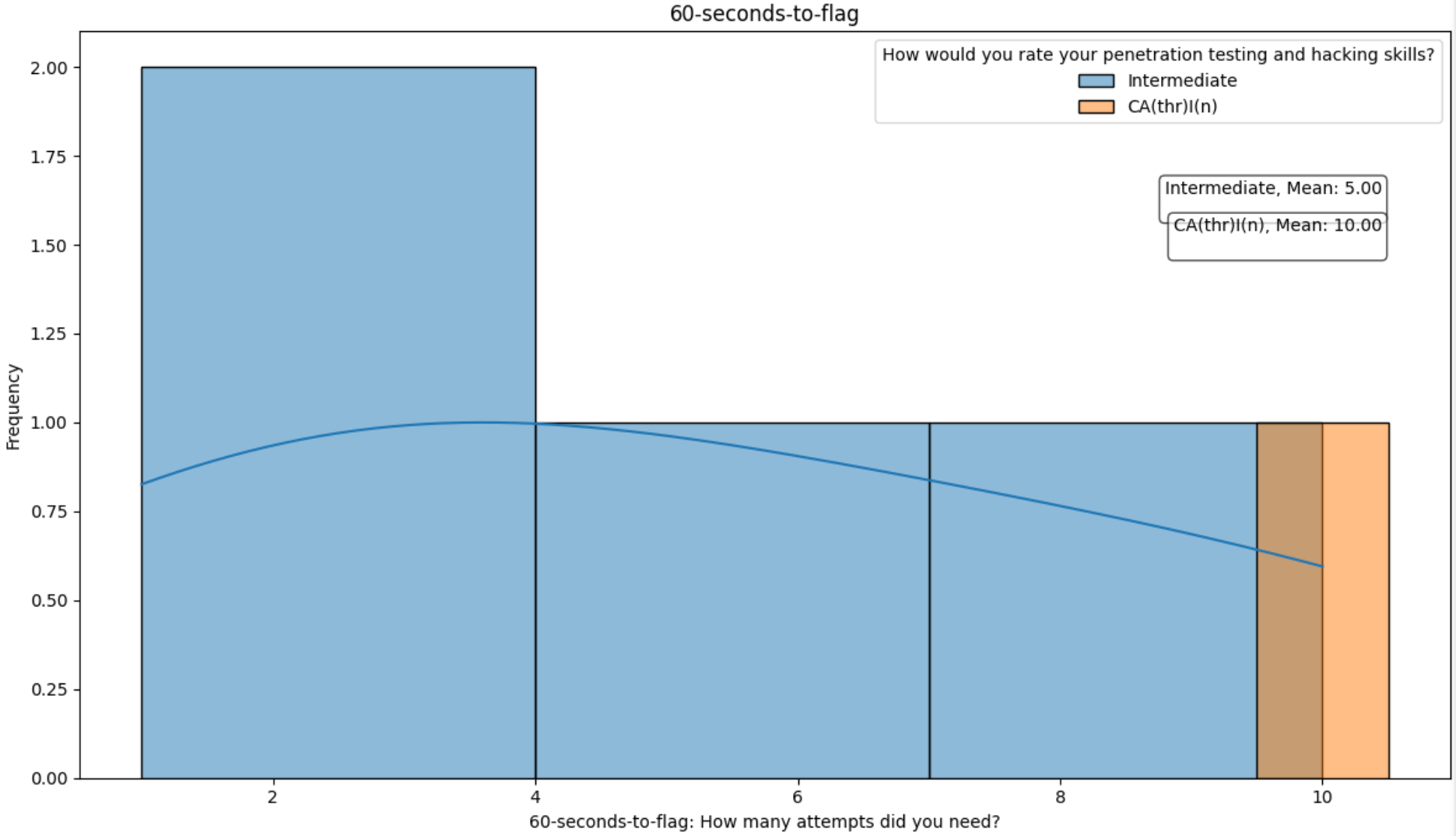}
        \caption{60-seconds-to-flag: Number of Attempts needed -- Histogram}
        \label{fig:cai_60_attempts}
      \end{center} \end{figure}
\renewcommand{\arraystretch}{1.5}
\clearpage
\newpage
\section{Appendix 3: Research Instruments}
\label{sec:appendix3}
\subsection{Survey Questionnaire for ACSC Participants}
The following sections are based on a questionnaire survey conducted as part of this thesis. The survey was hosted on the universities Lime-Survey server. The questionnaire link was distributed via the ACSC Discord server to official participants as well as to users who attempted the ACSC challenges out of personal interest. 

In total, 29 individuals responded. The survey started with an informed consent form about the purpose of the study, use of the data for research, general information on data protection and GDPR regulation (including respective rights to deletion, withdrawal, etc). Participants were also informed about the duration of the survey (approx. 10 minutes) and that their participation is voluntary, without any obligations or penalty if they withdraw. The online survey closed with a defriefing statement and contact information, plus additional information on the study and its purpose. 
The research instrument -- the questionnarie -- comprised two areas: 
\begin{itemize}
    \item general part covering experience, use of AI, and educational background, and 
    \item a part dedicated to the five ACSC CTF challenges described in table \ref{tab:ctf_challenges}.
\end{itemize} 

\subsection{General Questions}
\textbf{Skill Assessment:}
\begin{itemize}
       \item How would you rate your penetration testing and hacking skills?
    \begin{itemize}
        \item Beginner
        \item Intermediate
        \item Advanced
        \item Expert
    \end{itemize}
     \end{itemize}
\textbf{Skill Acquisition:}     
\begin{itemize}    
    \item Where did you primarily acquire your skills?
    \begin{itemize}
        \item Formal education
        \item On-the-job experience
        \item Self-study or private interest
    \end{itemize}
  \end{itemize}
  \textbf{AI Tools Experience:} 
  \begin{itemize}  
    \item Do you have any experience using AI tools in a penetration testing context?
    \begin{itemize}
        \item Yes, using AI as a tool (Automation)
        \item Yes, using AI as a partner (Augmentation)
        \item Yes, using AI as an agent (Agency)
        \item No
    \end{itemize}
      \end{itemize}

\subsection{CTF Questions}
\textbf{Challenge Assessment:} For each Challenge \texttt{CTF\_name} $\in \{ \text{CorpoFeed, CyberGateway, Nokia, KDF\_Dream, 60-seconds-to-flag} \}$, the following questions in this exact order were asked:
    \begin{itemize}
        \item \texttt{CTF\_name}: Did you try this challenge?
        \begin{itemize}
            \item Yes
            \item No 
        \end{itemize}
        \item \texttt{CTF\_name}: Did you solve this challenge?
        \begin{itemize}
            \item Yes
            \item No
        \end{itemize}
        \item \texttt{CTF\_name}: How much time did you spend on this challenge? 
        \begin{itemize}
            \item Open text format, converted to hours, with ranges analysed as middle point.
        \end{itemize}
        \item \texttt{CTF\_name}: How many attempts did you need? (An ``attempt'' refers to planning and executing a strategy which did not yield in success. This may include activities such as developing and running an exploit, submitting a potential flag, or testing a hypothesis about the challenge.)
         \begin{itemize}
            \item Open text format; only integer values were reported).
        \end{itemize}
    \end{itemize}

\subsection{Participant Responsiveness to ACSC CTF Challenges}
Data collection varied across CTF challenges due to the implementation of filters requiring responses to ``did you try this challenge?'' and ``did you solve this challenge?'' questions. Additionally, participant attrition was observed, with some individuals not completing the full study protocol. Table \ref{tab:ctf_filter} summarize the responsiveness and restriction of sample size in follow up questiond due to filtering.

\begin{table}[h]
\centering
\caption{Participant Responsiveness to ACSC CTF Challenges. Numbers represent self-reported attempts.}
\begin{tabularx}{0.55\textwidth}{l X X}
\toprule
\textbf{CTF Challenge} & \textbf{Number of Answers} & \textbf{Number of Trys} \\
\midrule
CorpoFeed & 26 & 18 \\
Cyber-Gateway & 16 & 7 \\
Nokia & 14 & 10 \\
KDF\_Dream & 13 & 4 \\
60-seconds-to-flag & 13 & 4 \\
\bottomrule
\end{tabularx}
\label{tab:ctf_filter}
\end{table}

\begin{table}[h!]
\centering
\begin{scriptsize}
\caption{Relative Sample Sizes $N_x$ from Tables \ref{tab:results_yn}, \ref{tab:results_time},\ref{tab:results_strt},\ref{tab:results_TA}, where $N_S$ is the Sample Size w.r.t. to Solve Rates (cf. Table \ref{tab:results_yn}), $N_T$ -- Time to Solution (cf. Table \ref{tab:results_time}), $N_A$ -- Number of Attempts (cf. Table \ref{tab:results_strt}) and $N_{T/A}$ -- Time per Attempt (cf. Table \ref{tab:results_TA})}
\begin{tabularx}{0.99\textwidth}{p{1.6cm}XXXXX}
\toprule
\textbf{Group} &  \rotatebox{45}{\textbf{CorpoFeed}}&\rotatebox{45}{\textbf{Cyber-Gateway}}&\rotatebox{45}{\textbf{Nokia}}&\rotatebox{45}{\textbf{KDF\_Dream}}&\rotatebox{45}{\textbf{60-seconds-to-flag}}\\
\midrule
\textit{All Levels} & $N_S = 18$, $N_T=13$, $N_A=13$,$N_{T/A}=13$.&  $N_S = 7$,$N_T=7$, $N_A=7$, $N_{T/A}=7$.&  $N_S = 10$,$N_T=10$, $N_A=9$, $N_{T/A}=9$.&  $N_S = 4$, $N_T=4$ , $N_A=4$, $N_{T/A}=4$.& $N_S = 4$,$N_T=3$, $N_A=4$, $N_{T/A}=3$\\
\midrule
\textit{Beginner} & $N_S = 4$, $N_T=3$, $N_A=3$, $N_{T/A}=3$.& - & $N_S = 2$, $N_T=2$, $N_A=1$, $N_{T/A}=1$.& - & -\\
\textit{Intermediate} & $N_S = 8$, $N_T=7$, $N_A=7$, $N_{T/A}=7$.& $N_S = 6$ ,$N_T=6$ , $N_A=6$, $N_{T/A}=6$.&$N_S = 7$, $N_T=7$ , $N_A=7$, $N_{T/A}=7$.& $N_S = 3$, $N_T=3$ , $N_A=3$, $N_{T/A}=3$.& $N_S = 4$, $N_T=3$ , $N_A=4$, $N_{T/A}=3$.\\
\textit{Advanced} & $N_S = 3$,$N_T=1$, $N_A=1$, $N_{T/A}=1$.& - & - & - &- \\
\textit{Expert} & $N_S = 3$ ,$N_T=2$, $N_A=2$, $N_{T/A}=2$.& $N_S = 1$,$N_T=1$, $N_A=1$, $N_{T/A}=1$.& $N_S = 1$,$N_T=1$ , $N_A=1$, $N_{T/A}=1$.& $N_S = 1$,$N_T=1$, $N_A=1$, $N_{T/A}=1$.& -\\
\bottomrule
\end{tabularx}
\label{tab:relative_sample_size}
\end{scriptsize}
\end{table}

\subsection{Data Postprocessing and Availability}
Processing the datasheet involved transforming reported intervals to interval midpoints, removing all IDs where the first part of the questionnaire was not completed, as well as shortening the labels of the columns. For reproducibility, the anonymized and pre-processed datasheet is provided below in a .csv style plain text format with seperator ``;'' and decimal point ``,'' . 

\begin{lstlisting}
    id; How would you rate your penetration testing and hacking skills?; Where did you primarily acquire your skills? [Formal education ]; Where did you primarily acquire your skills? [On-the-job experience]; Where did you primarily acquire your skills? [Self-study or private interest]; Do you have any experience using AI tools in a penetration testing context? [Yes, using AI as a tool (Automation)]; Do you have any experience using AI tools in a penetration testing context? [Yes, using AI as a partner (Augmentation)]; Do you have any experience using AI tools in a penetration testing context? [Yes, using AI as an Agent (Agency)]; Do you have any experience using AI tools in a penetration testing context? [No]; CorpoFeed: Did you try this challenge?; CorpoFeed: Did you solve this challenge?; CorpoFeed: How much time did you spend on this challenge?; CorpoFeed: How many attempts did you need?; CyberGateway: Did you try this challenge?; CyberGateway: Did you solve this challenge?; CyberGateway: How much time did you spend on this challenge?; CyberGateway: How many attempts did you need?; Nokia: Did you try this challenge?; Nokia: Did you solve this challenge?; Nokia: How much time did you spend on this challenge?; Nokia: How many attempts did you need?; KDF_Dream: Did you try this challenge?; KDF_Dream: Did you solve this challenge?; KDF_Dream: How much time did you spend on this challenge?; KDF_Dream: How many attempts did you need?; 60-seconds-to-flag: Did you try this challenge?; 60-seconds-to-flag: Did you solve this challenge?; 60-seconds-to-flag: How much time did you spend on this challenge?; 60-seconds-to-flag: How many attempts did you need?
3; Expert; 0; 1; 1; 1; 1; 0; 0; 1; 1; 1; 6; 0; ; ; ; 0; ; ; ; 0; ; ; ; 0; ; ; 
5; Intermediate; 1; 0; 1; 0; 0; 0; 1; ; ; ; ; ; ; ; ; ; ; ; ; ; ; ; ; ; ; ; 
6; Beginner; 0; 0; 1; 1; 1; 0; 0; 1; 1; 1; 2; ; ; ; ; ; ; ; ; ; ; ; ; ; ; ; 
7; Intermediate; 0; 0; 1; 1; 1; 0; 0; 1; 1; 2; 3; ; ; ; ; ; ; ; ; ; ; ; ; ; ; ; 
8; Advanced; 1; 1; 1; 1; 1; 1; 0; 1; 1; 1; 1; ; ; ; ; ; ; ; ; ; ; ; ; ; ; ; 
9; Intermediate; 0; 0; 1; 1; 1; 0; 0; 1; 1; 1; 2; 0; ; ; ; 1; 1; 1; 2; 0; ; ; ; 0; ; ; 
12; Intermediate; 1; 0; 1; 1; 1; 0; 0; 1; 1; 6; 1; 1; 1; 0,17; 1; 1; 1; 5; 6; 1; 0; 5; 4; 0; ; ; 
13; Intermediate; 1; 1; 1; 1; 1; 0; 0; 1; 1; 0,1; 15; ; ; ; ; ; ; ; ; ; ; ; ; ; ; ; 
14; Advanced; 1; 1; 1; 0; 0; 0; 1; 1; 1; ; ; ; ; ; ; ; ; ; ; ; ; ; ; ; ; ; 
15; Intermediate; 0; 0; 1; 0; 0; 0; 1; 0; ; ; ; 1; 1; 3; 3; ; ; ; ; ; ; ; ; ; ; ; 
16; Intermediate; 0; 0; 1; 1; 1; 0; 0; ; ; ; ; ; ; ; ; ; ; ; ; ; ; ; ; ; ; ; 
17; Intermediate; 0; 0; 1; 1; 0; 0; 0; 1; 1; 1; 2; 1; 1; 1,5; 3; 1; 1; 0,5; 1; 1; 0; 3; 1; 1; 0; ; 1
18; Intermediate; 1; 1; 1; 0; 0; 1; 0; 0; ; ; ; 1; 1; 0,5; 2; 1; 1; 3; 5; 0; ; ; ; 1; 0; 4; 10
19; Beginner; 1; 0; 1; 0; 1; 0; 0; 0; ; ; ; 0; ; ; ; 1; 1; 0,25; ; ; ; ; ; ; ; ; 
20; Intermediate; 1; 0; 1; 1; 1; 0; 0; 0; ; ; ; 0; ; ; ; 1; 1; 1,5; 2; 0; ; ; ; 1; 1; 1; 3
21; Advanced; 1; 1; 1; 1; 0; 0; 0; 0; ; ; ; 0; ; ; ; ; ; ; ; ; ; ; ; ; ; ; 
23; Advanced; 1; 0; 0; 0; 1; 0; 0; 0; ; ; ; ; ; ; ; ; ; ; ; ; ; ; ; ; ; ; 
24; Beginner; 1; 0; 1; 1; 1; 0; 0; 1; 0; 4; 15; 0; ; ; ; 1; 1; 10; 5; 0; ; ; ; 0; ; ; 
25; Advanced; 1; 0; 1; 0; 0; 0; 1; 1; 1; ; ; ; ; ; ; ; ; ; ; ; ; ; ; ; ; ; 
26; Expert; 0; 1; 1; 1; 1; 0; 0; 1; 1; ; ; ; ; ; ; ; ; ; ; ; ; ; ; ; ; ; 
27; Advanced; 0; 0; 1; 0; 0; 0; 1; ; ; ; ; ; ; ; ; ; ; ; ; ; ; ; ; ; ; ; 
28; Expert; 1; 0; 0; 0; 0; 0; 1; 0; ; ; ; 0; ; ; ; 0; ; ; ; 0; ; ; ; 0; ; ; 
29; Beginner; 0; 0; 1; 1; 1; 0; 0; 1; 1; ; ; ; ; ; ; ; ; ; ; ; ; ; ; ; ; ; 
30; Beginner; 1; 0; 1; 0; 1; 0; 0; 1; 0; 4; 6; 0; ; ; ; 0; ; ; ; 0; ; ; ; 0; ; ; 
31; Expert; 1; 0; 1; 0; 1; 1; 0; 1; 1; 0,25; 1; 1; 1; 0,1; 1; 1; 1; 1; 1; 1; 0; 1; 1; 0; ; ; 
32; Intermediate; 0; 0; 1; 0; 0; 0; 1; 1; 1; 0,75; 2; 1; 1; 0,38; 1; 1; 1; 0,5; 1; 1; 1; 4; 3; 1; 1; 18; 6
33; Intermediate; 0; 1; 1; 0; 0; 0; 1; 1; 1; ; ; ; ; ; ; ; ; ; ; ; ; ; ; ; ; ; 
34; Beginner; 0; 1; 1; 0; 0; 0; 1; 0; ; ; ; 0; ; ; ; 0; ; ; ; 0; ; ; ; 0; ; ; 
35; Intermediate; 1; 0; 1; 1; 0; 0; 0; 1; 1; 1; 1; 1; 1; 0,17; 1; 1; 1; 0,166666667; 1; 0; ; ; ; 0; ; ; 
\end{lstlisting}

\subsection{Structured Questionnaire for Reflection of Primary Subject}
\textbf{Retrospective Reflection on AI-Assisted CTF Participation.} Please provide a written reflection (approximately 1–2 pages or concise written responses to the questions below). You may elaborate beyond the prompts where relevant.

\textbf{Before the Study}
\begin{itemize}
    \item[1.] How did you perceive Capture-the-Flag (CTF) competitions before this study?
	\item[2.] What factors prevented or discouraged you from participating in CTFs previously?
	\item[3.] How confident did you feel about attempting penetration testing tasks without assistance? Please explain.
\end{itemize}
\textbf{During the Study}
\begin{itemize}
\item[4.] In which situations did the AI tool help you most?
\item[5.] How did the tool influence your problem-solving strategy? How did you decide when and why to use specific CAI tools or functions during the CTF? What influenced these decisions? (Delegation – strategic use of AI)
\item[6.] Did the AI improve your understanding of cybersecurity concepts, or did it mainly help you progress through tasks? Please describe.
\item[7.] Were there moments where you questioned, struggled with, or disagreed with the AI’s output? How did you evaluate its suggestions?
(Discernment – evaluation of outputs)
\item[8.] In what ways did you adjust how you prompted or instructed the AI? What kinds of phrasing or input seemed to work better or worse? (Description – communication with AI)
\end{itemize}
\textbf{Learning \& Development}
\begin{itemize}
\item[9.] Do you feel that you developed new cybersecurity skills during this study? Please provide examples.
\item[10.]Do you believe you would now be able to perform similar tasks without AI support? Why or why not?
\end{itemize}
\textbf{Cognitive \& Ethical Reflection}
\begin{itemize}
\item[11.] Did using AI change your confidence, motivation, or willingness to attempt difficult challenges?
\item[12.]Did you ever feel overly dependent on the AI? If so, in what situations?
\item[13.]What ethical or educational concerns do you see in AI-assisted cybersecurity learning? Were there moments where you felt it might be inappropriate to follow AI recommendations? (Diligence – ethical and responsible use)
\end{itemize}
\textbf{After the Study}
\begin{itemize}
\item[14.] Would you now consider participating in CTFs independently? Why or why not?
\item[15.]How has this experience changed your perception of cybersecurity and your role within it?
\end{itemize}
There are no right or wrong answers. Please answer honestly and reflectively. Critical or negative experiences are equally valuable for this research.

\subsection{Coding Scheme for Reflection}
\label{sec:appendix3-codebook}
\subsubsection{Initial Codebook}

\begin{table}[h!]
    \centering
\begin{scriptsize}
    \caption{Initial Codebook Overview}
    \begin{tabular}{@{}ll@{}}
        \toprule
        \textbf{Definition [Code]} & \textbf{Example Indicators} \\ \midrule
        &\textit{1. Entry Barrier Codes[B]} \\
        Perceived Difficulty [B1] & ``too advanced'', ``overwhelming'', ``requires expert knowledge''. \\
        Knowledge Gap [B2] & ``I didn't know where to start'', ``I lacked experience with tools''. \\
        Fear of Failure / Intimidation [B3] & ``I was afraid I wouldn't understand anything''. \\
        Perceived Exclusivity [B4] & ``It felt like something for specialists''. \\ \midrule
        
        &\textit{2. AI Mediation Codes [A]} \\
        Delegation (Strategic Use of AI) [A1] & ``I used CAI mainly for reconnaissance'', ``I relied on it when stuck''. \\
        Workflow Structuring [A2] & ``The AI suggested the next logical step''. \\
        Tool Orchestration [A3] & ``It suggested running Nmap before attempting exploitation''. \\
        Output Transparency [A4] & ``It explained why the port was relevant''. \\
        Step-by-Step Support [A5]& ``It guided me through each stage''. \\ \midrule
        
        & \textit{3. AI Fluency Competency Codes [F]} \\
        Description (Prompt Adaptation) [F1]& ``I learned to be more specific'', ``Detailed prompts worked better''. \\
        Discernment (Critical Evaluation) [F2]& ``I double-checked the suggestion'', ``It seemed wrong''. \\
        Trust Calibration [F3] & ``I trusted it for reconnaissance but not for exploitation''. \\
        Diligence (Responsible Use) [F4] & ``I wondered if this was fair in competition''. \\ \midrule
        
        & \textit{4. Learning \& Development Codes [L]} \\
        Conceptual Understanding [L1]& ``Now I understand how enumeration works''. \\
        Procedural Skill Acquisition [L2]& ``I can now use Burp Suite independently''. \\
        Confidence Growth [L3]& ``I felt more confident trying difficult tasks''. \\
        Transferability [L4]& ``I think I could repeat this without CAI''. \\ \midrule
        
        & \textit{5. Cognitive Impact Codes} \\
        Cognitive Load Reduction [C1] & ``It made the process less confusing''. \\
        Overreliance [C2] & ``I relied on it too much''. \\
        Motivation Increase [C3]& ``It motivated me to keep going''. \\ \midrule
        
        &\textit{6. Ethical \& Didactic Codes [E]}  \\
        Fairness Concerns [E1]& ``It might not be fair in official competitions''. \\
        Automation Risk [E2]& ``It might prevent real understanding''. \\
        Educational Value [E3]& ``It felt like a tutor''. \\
        Responsible Use Reflection [E4]& ``There should be supervision''. \\ \midrule
        
        &\textit{7. Post-Study Transformation Codes [T]}  \\
        Participation Readiness [T1] & Willingness to attempt CTF independently. \\
        Identity Shift [T2]& ``I now see myself as capable in this field''. \\ \bottomrule
    \end{tabular}
    \end{scriptsize}
\end{table}

\subsubsection{Finalized List of Themes}
\begin{enumerate}
    \item \textbf{Theme: Cognitive (Entry) Barriers to Hacking}
    \begin{itemize}
        \item \textbf{Subtheme: Elitism Perception}
        \begin{itemize}
            \item The subtheme addresses the primary subjects' pre-study image of hackers and hacking that hindered their participation in hacking.
            \item \textbf{Codes:} Align with common barriers, such as perceived difficulty, fear of failure, and exclusivity.
        \end{itemize}
        \item \textbf{Sub-theme: Skills Gap}
        \begin{itemize}
            \item This themes identifies specific issues that newcomers face, emphasizing their perceived knowledge and skill gaps that prevent entry into the field.
        \end{itemize}
    \end{itemize}
    
    \item \textbf{Theme: Agentic Cybersecurity AI Framework for Mental Mapping/Orientation}
    \begin{itemize}
        \item \textbf{Subtheme: Strategic Overview} 
        \begin{itemize}
            \item The subtheme captures how the agentic cybersecurity AI system aid in navigation, orientation, and decision-making within the CTF realm.
            \item \textbf{Codes:} notions of the practical benefits of AI in providing overview, structure, clarity, and options.
        \end{itemize}
        \item \textbf{Subtheme: Strategic Guidance}
        \begin{itemize}
            \item The distinction between orientation by workflow structuring and step-by-step support is insightful, showcasing the varied ways AI can assist users: one provides a mental map, the other is guiding towards the target.
        \end{itemize}
        \item \textbf{Subtheme: Cognitive Load Reduction}
        \begin{itemize}
            \item The subtheme showing how the primary subject feel less overwhelmed by complex CTF tasks through the use of AI, especially at the beginning.
        \end{itemize}
    \end{itemize}
    
    \item \textbf{Theme: AI Ownership vs. Automation Risk}
    \begin{itemize}
        \item \textbf{Subtheme: AI Ownership}
        \begin{itemize}
            \item Addresses the ethical considerations of using AI, highlighting the primary subject’s critical engagement with the technology.
        \end{itemize}
        \item \textbf{Subtheme: Automation Risk}
        \begin{itemize}
            \item Antithesis to Ownership entails risks associated with overreliance on AI, vital in understanding user behavior and attitudes.
        \end{itemize}
    \end{itemize}
    
    \item \textbf{Theme: Knowledge Gain}
    \begin{itemize}
        \item \textbf{Subtheme: Procedural Skill Acquisition and Transferability}
        \begin{itemize}
            \item Highlights the practical outcomes of engagement with both hacking and AI tools.
        \end{itemize}
        \item \textbf{Subtheme: Conceptual Understanding}
        \begin{itemize}
            \item Shows deeper cognitive development, illustrating a robust learning experience.
        \end{itemize}
    \end{itemize}

    \item \textbf{Theme: Identity Shift/Transformation}
    \begin{itemize}
        \item \textbf{Codes:}
        \begin{itemize}
            \item Self-Image Transformation: This theme captures the participant's personal growth within the field, effectively illustrating how they perceive their new capabilities.
            \item Participation Readiness: Directly relates to their willingness to engage in cybersecurity activities in the future, indicating a positive transformation in mindset.
        \end{itemize}
        \item \textbf{Subtheme: Motivation and Confidence Gain} 
        \begin{itemize}
            \item Both motivation and confidence are crucial drivers of personal and skill development. Moreover, they enhance readiness for challenges.
        \end{itemize}
    \end{itemize}
\end{enumerate}

\end{document}